\documentclass[lettersize,journal]{IEEEtran}
\usepackage{amsmath,amsfonts}
\usepackage{subcaption}
\usepackage{algorithmic}
\usepackage[ruled,linesnumbered]{algorithm2e}
\usepackage{array}

\usepackage{textcomp}
\usepackage{stfloats}
\usepackage{url}
\usepackage{verbatim}
\usepackage{multirow}
\usepackage[normalem]{ulem}
\useunder{\uline}{\ul}{}
\usepackage{graphicx}
\usepackage{cite}
\usepackage{makecell}
\usepackage{enumitem}
\usepackage{graphicx}
\usepackage{float}
\usepackage{tcolorbox}
\usepackage{lineno,hyperref}
\usepackage{booktabs}
\usepackage{orcidlink}
\begin{document}

\title{From Collapse to Stability: A Knowledge-Driven Ensemble Framework for Scaling Up Click-Through Rate Prediction Models}

\author{Honghao Li\orcidlink{0009-0000-6818-7834}, \textit{Student Member, IEEE}, Lei Sang\orcidlink{0009-0007-1480-6522}, Yi Zhang\orcidlink{0000-0001-8196-0668}, Guangming Cui, and Yiwen Zhang*\orcidlink{0000-0001-8709-1088} 

\IEEEcompsocitemizethanks{\IEEEcompsocthanksitem Honghao Li, Lei Sang, Yi Zhang, and Yiwen Zhang are with the School of Computer Science and Technology, Anhui University 230601, Hefei, Anhui, China. (E-mail: salmon1802li@gmail.com, sanglei@ahu.edu.cn, zhangyi.ahu@gmail.com, and zhangyiwen@ahu.edu.cn)
\IEEEcompsocthanksitem Guangming Cui is with the School of Software, Nanjing University of Information Science \& Technology, Jiangsu Province Engineering Research Center of Advanced Computing and Intelligent Services, and State Key Laboratory for Novel Software Technology, Nanjing University, P.R. China. \protect~ (E-mail: gcui@nuist.edu.cn)
}
\thanks{*Corresponding author.}}

\markboth{Journal of \LaTeX\ Class Files,~Vol.~x, No.~x, August~xxxx}%
{Shell \MakeLowercase{\textit{et al.}}: From Collapse to Stability: A Study on Network Collapse in Scaling Up Recommendation Models}


\maketitle
\begin{abstract}
    Click-through rate (CTR) prediction plays a crucial role in modern recommender systems. While many existing methods utilize ensemble networks to improve CTR model performance, they typically restrict the ensemble to only two or three sub-networks. Whether increasing the number of sub-networks consistently enhances CTR model performance to align with scaling laws remains unclear. In this paper, we investigate larger ensemble networks and find three inherent limitations in commonly used ensemble methods: (1) performance degradation as the number of sub-networks increases; (2) sharp declines and high variance in sub-network performance; and (3) significant discrepancies between sub-network and ensemble predictions. Meanwhile, we analyze the underlying causes of these limitations from the perspective of dimensional collapse: the collapse within sub-networks becomes increasingly severe as the number of sub-networks grows, leading to a lower knowledge abundance.
  
   In this paper, we employ knowledge transfer methods, such as Knowledge Distillation (KD) and Deep Mutual Learning (DML), to address the aforementioned limitations. We find that KD enables CTR models to better follow scaling laws, while DML reduces variance among sub-networks and minimizes discrepancies with ensemble predictions. Furthermore, by combining KD and DML, we propose a model-agnostic and hyperparameter-free Knowledge-Driven Ensemble Framework (KDEF) for CTR Prediction. Specifically, we employ students' collective decision-making as an abstract teacher to guide each student (sub-network). Moreover, we encourage mutual learning among students to enable knowledge acquisition from different views. To address the issue of balancing the loss hyperparameters, we design a novel examination mechanism to ensure tailored teaching from teacher-to-student and selective learning in peer-to-peer. Experimental results on five real-world datasets demonstrate the effectiveness, compatibility, and flexibility of KDEF. The code, running logs, and detailed hyperparameter configurations are available at: \url{https://github.com/salmon1802/KDEF}.
\end{abstract}

\begin{IEEEkeywords}
Knowledge Distillation, Deep Mutual Learning, Ensemble Network, Recommender Systems, CTR Prediction.
\end{IEEEkeywords}

\section{Introduction}
\IEEEPARstart{C}{lick-through} rate (CTR) prediction is a cornerstone task in modern recommender systems \cite{Bars, TKDE1, TKDE2, aim, TKDE3}, aiming to estimate the likelihood that a user will click on a recommended item, such as an advertisement, product, or piece of content \cite{openbenchmark}. This task underpins numerous real-world applications, from online advertising and E-commerce to content personalization, where accurate predictions directly influence user engagement and revenue generation \cite{finalmlp, dcnv2, Bars}.

\begin{figure}[t]
    \setlength{\abovecaptionskip}{1pt}
    \setlength{\belowcaptionskip}{1pt}
\centering
    \begin{minipage}[t]{0.48\linewidth}
        \centering
        \includegraphics[width=\linewidth]{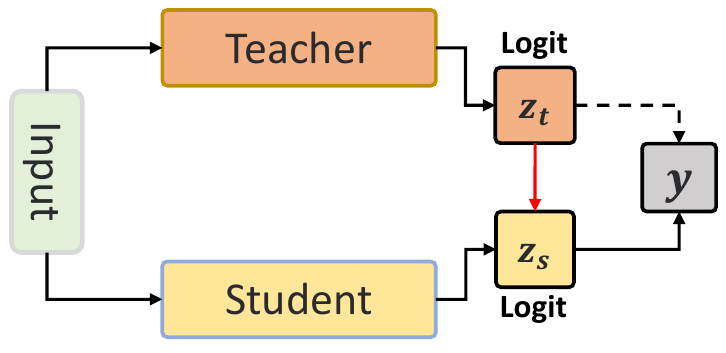}
        \subcaption{\footnotesize KD}
    \end{minipage}
    \hspace{-0.5em} 
    \begin{minipage}[t]{0.48\linewidth}
        \centering
        \includegraphics[width=\linewidth]{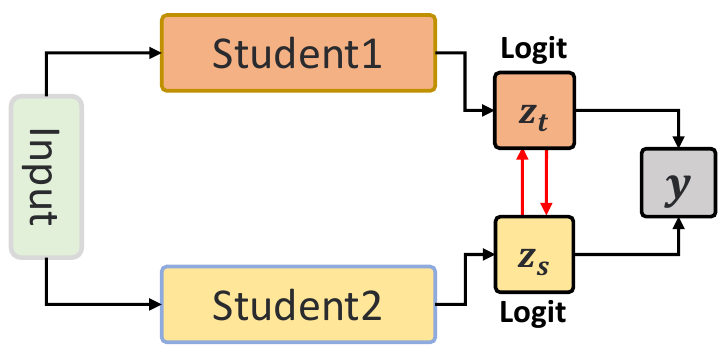}
        \subcaption{\footnotesize DML}
    \end{minipage} \\ 
    \begin{minipage}[t]{0.42\linewidth}
        \centering
        \includegraphics[width=\linewidth]{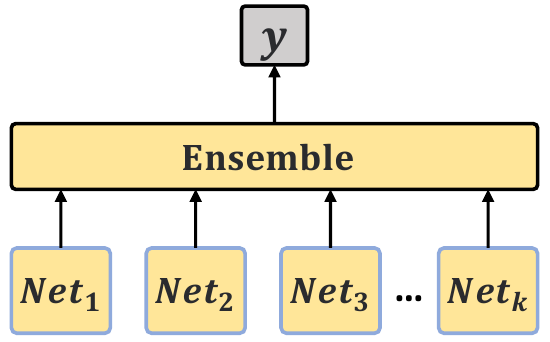}
        \subcaption{\footnotesize Ensemble Network for CTR Prediction}
    \end{minipage}
    \hspace{-0.5em} 
    \begin{minipage}[t]{0.48\linewidth}
        \centering
        \includegraphics[width=\linewidth]{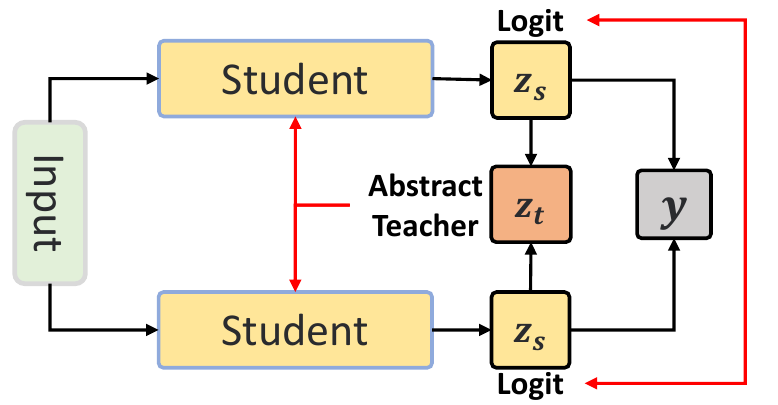}
        \subcaption{\footnotesize KDEF}
    \end{minipage}
    \captionsetup{justification=raggedright}
    \caption{Comparison of knowledge distillation, deep mutual learning, and ensemble methods.}
    \label{moti}
    \vspace{-1em}
\end{figure}

In CTR prediction tasks, ensemble\footnote{In this paper, we use the term “ensemble” to refer to the philosophy of integrating multiple sub-networks using end-to-end learning, as it is commonly adopted in the literature \cite{ensemble_first, GDCN, EulerNet, Wukong}. Some studies \cite{xdeepfm, dcnv2, autoint} also interpret it as a form of information fusion or network combination.} network \cite{ensemble} is an effective strategy for performance improvement \cite{deepfm, xdeepfm, dcnv2, KD4CTR}. Therefore, most CTR models proposed by researchers in both academia and industry aim to combine multiple \textit{parallel and independent} sub-networks \cite{finalmlp, FINAL, CETN, TF4CTR}, as shown in Fig. \ref{moti} (c). For instance, DHEN \cite{DHEN}, a deep and hierarchical ensemble model, along with its enhanced version on Pinterest \cite{PINTEREST}, exemplifies this approach. Similarly, HMoE \cite{tencent}, a feature multi-embedding and multi-tower architecture proposed by Tencent, and MBCnet \cite{MBCnet}, a multi-branch cooperation network proposed by Taobao. However, these model typically restrict the ensemble to only two or three sub-networks. Meanwhile, as the demand for higher performance drives the development of increasingly complex models, a critical question emerges: \textit{How does the model perform as the number of sub-networks increases?}

To answer this question, we empirically analyze the relationship between ensemble performance and the number of sub-networks, leading to three key findings (limitations):

\begin{itemize}[leftmargin=*]
\item Finding 1: \textbf{Performance degradation with more networks.} Some studies on scaling laws \cite{GPT3.0} suggest that in large language models, performance tends to improve as the number of parameters increases. However, we observe an opposite trend in CTR prediction task, where the more sub-networks are included in the ensemble, the lower the ensemble model performance becomes.
\item Finding 2: \textbf{Sharp decline and high variance in sub-networks performance.} We further investigate the performance trends of individual sub-networks and observe that their performance declines more sharply compared to the ensemble network as the number of sub-networks increases. Moreover, there is a significant variance between the best and worst-performing sub-networks.
\item Finding 3: \textbf{Large discrepancies between sub-network and ensemble prediction.} We observe a significant gap between the performance of individual sub-networks and the ensemble network, with this gap increasing as the number of sub-networks grows, even reaching as high as 6\%. This limits the model's flexibility in practice. Ideally, if we could train sub-networks to approximate the ensemble’s performance, we could deploy sub-networks when lower model complexity is required and use the ensemble when higher performance is needed.
\end{itemize}

Furthermore, we investigate the causes of these limitations from the perspective of dimensional collapse. As shown in Fig. \ref{moti} (c), each sub-network is trained independently with a 1-bit click signal, lacking knowledge exchange. Consequently, as the number of sub-networks increases, their Knowledge Abundance (KA) decreases.  Motivated by this observed deficiency in KA, we believe that the introduction of knowledge transfer methods between sub-networks may be effective in mitigating the issue of low KA \cite{KD_survey1, KD_survey2}. For example, Knowledge Distillation (KD) \cite{KD} can transfer knowledge from a teacher model to a smaller student model to improve its performance (Fig. \ref{moti} (a)). Deep Mutual Learning (DML) \cite{DML} replaces the teacher model with multiple student models that learn from each other, enabling peer-to-peer knowledge transfer (Fig. \ref{moti} (b)).

Therefore, we investigate how KD and DML can mitigate these limitations. For KD, we employ the collective decision-making of the student networks as an abstract teacher to guide the learning of each sub-network. Empirical results demonstrate that this KD-based ensemble approach mitigates the limitations outlined in Findings 1 \& 2. For DML, each sub-network receives supervision from the ground truth labels while being encouraged to learn from each other. Empirical results show that while DML does not address the limitation in Finding 1, it improves sub-network performance over KD, better addressing the limitations in Finding 2 \& 3. 

To address these limitations simultaneously, a simple idea is to combine KD and DML, with DML improving communication among students in KD. However, this approach faces practical challenges, such as tuning hyperparameters in a large search space of $O(K)$ (teacher supervising $K$ students) and $O(K(K-1))$ (students learning from each other). To solve this, we propose a novel loss adaptive balancing strategy, called the \textbf{examination mechanism}, which allows teachers to tailor instruction and students to selectively learn from each other, avoiding knowledge conflicts \cite{KDCL}.

Finally, we propose the \textbf{Knowledge-Driven ensemble Framework (KDEF)} for CTR prediction, with its architecture illustrated in Fig. \ref{moti} (d). KDEF abandons the approach of designing explicit teacher models and instead utilizes the collective decision-making of students as an abstract teacher. This facilitates the generation of high-quality soft labels to guide the learning of multiple student models (sub-networks), aiming to distill group knowledge into individual students. Meanwhile, KDEF also encourages mutual learning among students, helping each student model gain knowledge from different views. Through the examination mechanism, we adaptively assign weights to the distillation loss and mutual learning loss for each sub-network, enabling tailored teaching by the teacher and selective learning by peer-to-peer.

The core contributions of this paper are summarized as follows:
\begin{itemize}[leftmargin=*]
\item To the best of our knowledge, this is the first work to empirically identify three key limitations of ensemble network in CTR prediction and to explore effective solutions of limitations through end-to-end KD and DML.
\item We propose a novel model-agnostic and hyperparameter-free Knowledge-Driven Ensemble Framework to improve ensemble network and sub-network performance.
\item We design a novel examination mechanism that adaptively
assigns loss weights to ensure tailored teaching from teacher-to-student and selective learning in peer-to-peer.
\item We conduct comprehensive experiments on five real-world datasets to demonstrate the effectiveness, compatibility, and flexibility of the KDEF.
\end{itemize}

\section{Related Work}
\subsection{CTR Prediction}
Effectively capturing feature interactions is a crucial method for improving performance in CTR prediction tasks. Traditional CTR models, such as LR \cite{LR} and FM \cite{FM}, can only capture low-order feature interactions, which limits their performance. With the rise of deep learning, several works \cite{DNN, NFM, widedeep} have demonstrated that multi-layer perceptrons (MLP) can capture high-order feature interactions, leading to further performance improvements. As a result, recent CTR models \cite{FINAL, CETN, adagin, SimCEN} have begun ensemble different sub-networks to capture both low-order and high-order feature interactions simultaneously. For example, DeepFM \cite{deepfm}, DCNv2 \cite{dcnv2}, Wide \& Deep \cite{widedeep}, EDCN \cite{EDCN}, xDeepFM \cite{xdeepfm}, FinalMLP \cite{finalmlp}, GDCN \cite{GDCN}, FCN \cite{FCN} etc. In industrial recommender systems, such ensemble networks are also commonly used. Zhang \textit{et al.} \cite{DHEN} \cite{PINTEREST} perform hierarchical deep ensemble on various CTR models, while Pan \textit{et al.} \cite{tencent} integrate multiple CTR models using multi-embedding and multi-tower architectures. Chen \textit{et al.} \cite{MBCnet} propose a multi-branch cooperation network to enhance the collaboration among sub-networks. These ensemble models typically aggregate all sub-networks using summation, averaging, or concatenation, achieving certain performance improvements. However, most of the above studies lack exploration of the larger ensemble. The relationship between the performance of individual sub-networks and the overall model also remains unclear. Therefore, this paper aims to explore the limitations of larger ensemble models and address them through our proposed framework.

\subsection{Knowledge Transfer for CTR Prediction}
KD \cite{KD} and DML \cite{DML} aim to transfer knowledge between different networks to improve model performance \cite{KD_survey1, KD_survey2, DML}. Both methods have achieved significant advances in computer vision and natural language processing \cite{KDCL, TeacherKD1, TeacherKD2, DKD, KDNLP}. Recently, there have also been some developments in applying KD to CTR prediction. Zhu \textit{et al.} \cite{KD4CTR} first introduces KD into CTR prediction, using an ensemble network of teachers to enhance the performance of the student network. Tian \textit{et al.} \cite{DAGFM} distills information captured by DNN into a graph model, improving performance while maintaining high inference efficiency. Deng \textit{et al.} \cite{BKD} introduces a bridge model between the student and teacher to facilitate student learning. Liu \textit{et al.} \cite{PositionKD} transfers positional knowledge from the teacher model to the student model.

In contrast, there has been \textit{very limited} exploration of combining DML with CTR prediction. \cite{DML4CTR} is the first and, to date, the only work to introduce DML into CTR prediction, which uses DML to further fine-tune pre-trained models. However, this paper aims to explore how KD and DML can address the limitations of the ensemble network in an end-to-end manner.

\section{Preliminaries}
\subsection{CTR Prediction Task}
CTR prediction is typically considered a binary classification task that utilizes user profiles, item attributes, and context as features to predict the probability of a user clicking on an item \cite{openbenchmark, autoint}. The composition of these three types of features is as follows:
\begin{itemize} 
\item  \emph{User profiles} ($p$): age, gender, occupation, etc.
\item \emph{Item attributes} ($a$): brand, price, category, etc.
\item \emph{Context} ($c$): timestamp, device, position, etc.
\end{itemize}
Further, we can define a CTR sample in the tuple data format: $X = \{x_p, x_a, x_c\}$. Variable $y \in \{0, 1\}$ is an true label for user click behavior:
\begin{equation}
    y= \begin{cases}1, & \text{user} \text { has clicked } \text{item}, \\ 0, & \text {otherwise.}\end{cases}
\end{equation}
It is a positive sample when $y=1$ and a negative sample when $y=0$. The final purpose of the CTR prediction model, which is to reduce the gap between the model's prediction and the true label, is formulated as follows: 
\begin{equation}
    \hat{y} = \texttt{MODEL}(X; \Theta), \quad \Theta^{*} = \arg\min_{\Theta} \|y - \hat{y}\|
\end{equation}
where $\hat{y}$ is the final result of the model prediction, \texttt{MODEL} denotes the CTR model, and $\Theta^{*}$ denotes the optimal parameters of the model.

\subsection{Dimensional Collapse}
Dimensional collapse \cite{embedding_Collapse,Dimensional_Collapse} refers to the phenomenon in deep neural networks where feature representation diversity gradually diminishes, and information is compressed into a low-dimensional subspace. Specifically, when dimensional collapse occurs, the rank of the parameter matrix significantly decreases, exhibiting low-rank characteristics, which prevents the full utilization of the entire representation space. In self-supervised or contrastive learning \cite{feature_Collapse, Dimensional_Collapse}, the feature vectors produced by the encoder may collapse into a low-dimensional subspace or become nearly linearly dependent, making it difficult for the encoder to distinguish representations of different input samples. Therefore, this phenomenon severely degrades the performance of downstream tasks.

Jing \textit{et al.} \cite{Dimensional_Collapse} propose a metric to quantify the degree of collapse. Specifically, given a parameter matrix $\boldsymbol{W}$, we perform singular value decomposition \cite{SVD} (SVD) as $\boldsymbol{W} = \boldsymbol{U} \boldsymbol{\Sigma} \boldsymbol{V^\top}$, where $\boldsymbol{\Sigma}$ contains the singular values $\boldsymbol{\sigma}$ in ascending order. A parameter matrix without dimensional collapse should have non-zero singular values across all dimensions. In contrast, a greater number of near-zero singular values indicates a more severe degree of dimensional collapse. Meanwhile, larger singular values indicate more information (knowledge) in that dimension, and vice versa.


To better quantify the degree of collapse of the matrix $\boldsymbol{W}$ across different layers in the sub-network, we propose \textbf{knowledge abundance} (KA) as a quantification method: 
\begin{equation}  
\begin{aligned}
\mathrm{KA}(\boldsymbol{W}_l, \tau)=\frac{\sum_{i=1}^m \mathbb{I}(\boldsymbol{\sigma}_i > \tau)}{\sum_{i=1}^m \mathbb{I}(\boldsymbol{\sigma}_i > 0)},
\end{aligned}
\end{equation}
where $m$ is the number of singular values, and \(\mathbb{I}(\cdot)\) is the indicator function, defined as:
\begin{equation}
\mathbb{I}(\boldsymbol{\sigma}_i > \tau) = 
\begin{cases} 
1 & \text{if } \boldsymbol{\sigma}_i > \tau, \\
0 & \text{otherwise},
\end{cases}
\end{equation}
where $\boldsymbol{W}_l$ denotes the $l$-th layer matrix, \(\tau\) is a predefined threshold, \(d\) denotes the number of singular values of \( \boldsymbol{W} \), and \(\boldsymbol{\sigma}_i\) is the \(i\)-th singular value of \( \boldsymbol{W}_l \). This metric measures the proportion of significant singular values, reflecting the diversity and capacity of sub-network representations. Higher KA indicates reduced dimensional collapse, while lower KA suggests information compression into fewer dimensions.

\section{Optimizing Ensemble Network for CTR Prediction: From  Collapse to Stability}
\label{section 3}
In this section, we investigate the performance trends of large-scale ensemble networks in CTR prediction tasks. Through an empirical study, we identify three inherent limitations of such networks and analyze their underlying causes from the perspective of dimensional collapse. This insight motivates us to introduce knowledge transfer methods to alleviate these limitations. However, we observe that both KD-based and DML-based methods address only part of the limitations. To overcome this, we propose KDEF, a model-agnostic and hyperparameter-free framework that seamlessly integrates both methods while introducing an Examination Mechanism to adaptively balance multiple loss terms. This design facilitates more effective knowledge transfer and collaborative learning among sub-networks, ultimately stabilizing improves the overall performance of the ensemble network.
\subsection{Collapse Phenomenon for Large Ensemble Network in CTR Prediction}
\label{Collapse_Phenomenon}
\subsubsection{Empirical Analysis}
\begin{figure}[t]
\centering
    \begin{minipage}[t]{0.45\linewidth}
        \centering
        \includegraphics[width=\linewidth]{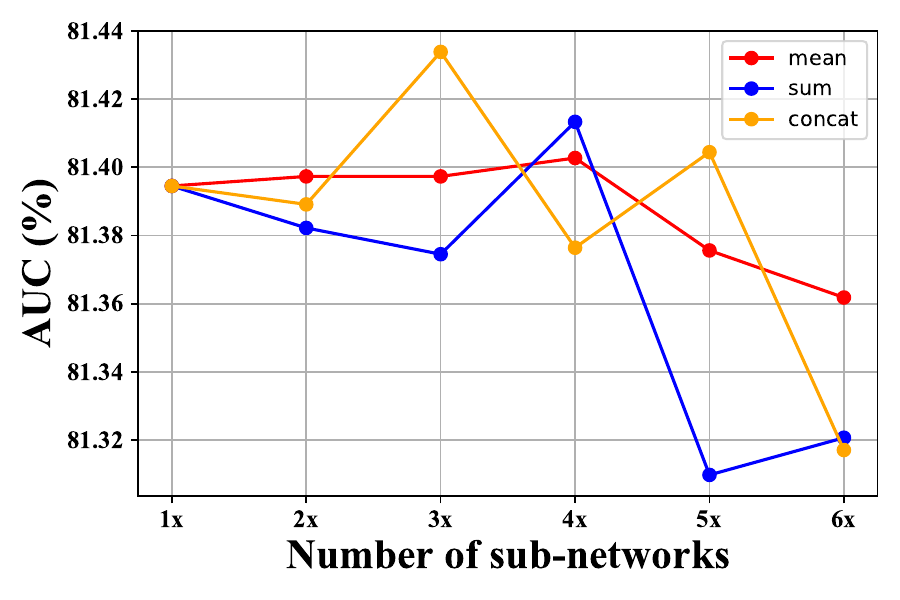}
        \subcaption{\footnotesize Ensemble prediction result $\hat{y}$}
    \end{minipage}
    \hspace{-0.5em} 
    \begin{minipage}[t]{0.45\linewidth}
        \centering
        \includegraphics[width=\linewidth]{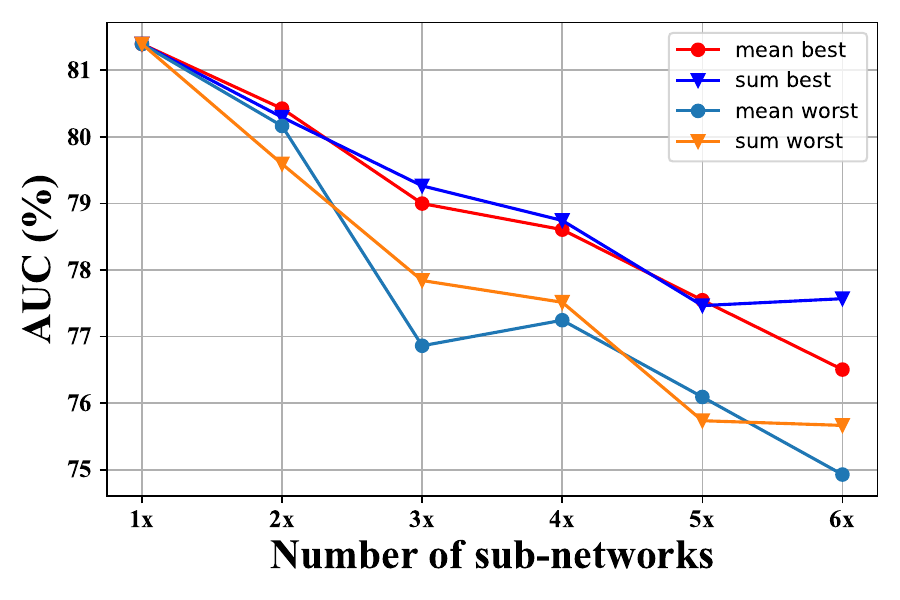}
        \subcaption{\footnotesize Best and worst sub-network.}
    \end{minipage}%
    \captionsetup{justification=raggedright}
    \caption{The performance changes of different ensemble methods on the well-known large-scale Criteo dataset \cite{Criteo}. Due to the inability of the concatenation method to compute sub-network predictions, it is omitted in subfigure (b).}
    \label{ensemble}
\vspace{-1em}
\end{figure}

Ensemble network \cite{KD4CTR} is a commonly used paradigm for performance enhancement in CTR prediction tasks \cite{finalmlp, TF4CTR}. Most CTR models employ simple operations such as summation \cite{deepfm,xdeepfm}, mean \cite{FINAL, FCN}, or concatenation \cite{dcn, dcnv2} to aggregate the predictions of multiple sub-networks, thereby improving model performance. The formalization of these ensemble methods is as follows:
\begin{equation}
\label{ensemble_prediction}
\begin{aligned}
\hat{y}_t &= \texttt{Ensemble}(\hat{y}_{s,1}, \hat{y}_{s,2}, \dots, \hat{y}_{s,k}),
\end{aligned}
\end{equation}
where $\hat{y}_t$ is the ensemble prediction result, $\hat{y}_{s,k}$ represents the prediction result of the $k$-th sub-network (student network), and \texttt{Ensemble} denotes different fusion functions. However, these ensemble methods often only fuse the predictions from two sub-networks \cite{deepfm, FCN, GDCN} and lack exploration into the scaling laws for more sub-networks. Therefore, we monitor the impact of the aforementioned three common fusion methods on the performance of different numbers of sub-networks, as shown in Fig. \ref{ensemble}. We have the following findings\footnote{To ensure a fair comparison, all experiments in Section \ref{section 3} are conducted using MLPs with the same architecture [400, 400, 400], which is a common architectural setup in CTR predictions \cite{GDCN, finalmlp, EulerNet}. Meanwhile, in this paper, we use '$\Delta \mathbf{x}$' to denote an ensemble network consisting of $\Delta$ sub-networks.}:

\begin{tcolorbox}[colframe=black!75!white, colback=white!95!blue, arc=3mm, boxrule=0.5mm, width=\linewidth]
\label{finding}
Finding 1 (Limitation). \textbf{Performance degradation with more networks}: From Fig. \ref{ensemble} (a), we observe that, regardless of the ensemble method used, the ensemble prediction performance of the model gradually decreases as the number of sub-networks increases. Furthermore, the sum and concat methods exhibit greater instability compared to the mean method.
\end{tcolorbox}

\begin{tcolorbox}[colframe=black!75!white, colback=white!95!blue, arc=3mm, boxrule=0.5mm, width=\linewidth]
Finding 2 (Limitation). \textbf{Sharp decline and high variance in sub-networks performance.}: As shown in Fig. \ref{ensemble} (b), the performance of the sub-network degrades more significantly as the number of sub-networks increases. Meanwhile, there exists a substantial variance in the performance of sub-networks.
\end{tcolorbox}

\begin{tcolorbox}[colframe=black!75!white, colback=white!95!blue, arc=3mm, boxrule=0.5mm, width=\linewidth]
Finding 3 (Limitation). \textbf{Large discrepancies between sub-network $\hat{y}_{s}$ and ensemble prediction $\hat{y}$}: Synthesizing the experimental results from Fig. \ref{ensemble}, we observe that even the best-performing sub-networks exhibit significant discrepancies compared to the ensemble prediction results in $\hat{y}$. For instance, with a 6x sub-network configuration, the performance gap between the mean method ensemble $\hat{y}$ and best $\hat{y}_{s}$ reaches 6\%. This means that we can only use the model after ensemble, which reduces model flexibility.
\end{tcolorbox}

These three findings confirm that further performance improvement cannot be achieved through conventional sub-network ensemble methods. However, some studies on scaling laws \cite{GPT3.0} suggest that model performance typically improves with an increase in the number of parameters, and may even exhibit emergent phenomena \cite{Emergent}. This motivates us to further investigate the reasons behind this contradiction with established conclusions in other research areas.

\subsubsection{Cause Analysis}
To further investigate the reason behind the performance degradation caused by increasing the number of sub-networks, we conduct a detailed analysis based on mean fusion, as it exhibits relatively smooth performance changes compared to other ensemble methods. As illustrated in Fig. \ref{Net_collapse_all}, we present a layer-wise visualization of the singular values for both the $1\mathbf{x}$ and $6\mathbf{x}$ ensemble networks. Based on these results, we have the following observations:
\begin{itemize}[leftmargin=*]
\item We observe that Layer 1, directly connected to feature embeddings, consistently exhibits relatively uniform singular values.
\item We observe that singular values progressively decrease across layers, as evidenced in Fig. \ref{Net_collapse_all} (b), where knowledge abundance (KA) drops from 32.25\% in the first layer to 6.75\%, indicating gradual compression of information into a low-rank matrix.
\item We observe that, as shown in Fig.~\ref{Net_collapse_all} (a) and (b), each layer of the $1\mathbf{x}$ ensemble networks exhibits higher KA compared to the $6\mathbf{x}$ ensemble networks. For instance, in Layer 2, the KA difference between the two reaches 4.75\%.
\end{itemize}
These observations suggest that a single network is able to better utilize the feature representation space, while introducing more sub-networks or deeper layers exacerbates the problem of dimensional collapse. Therefore, we conclude that:

\begin{figure*}[t]
\vspace{-1em}
\centering
    \begin{minipage}[t]{0.49\linewidth}
        \centering
        \includegraphics[width=\linewidth]{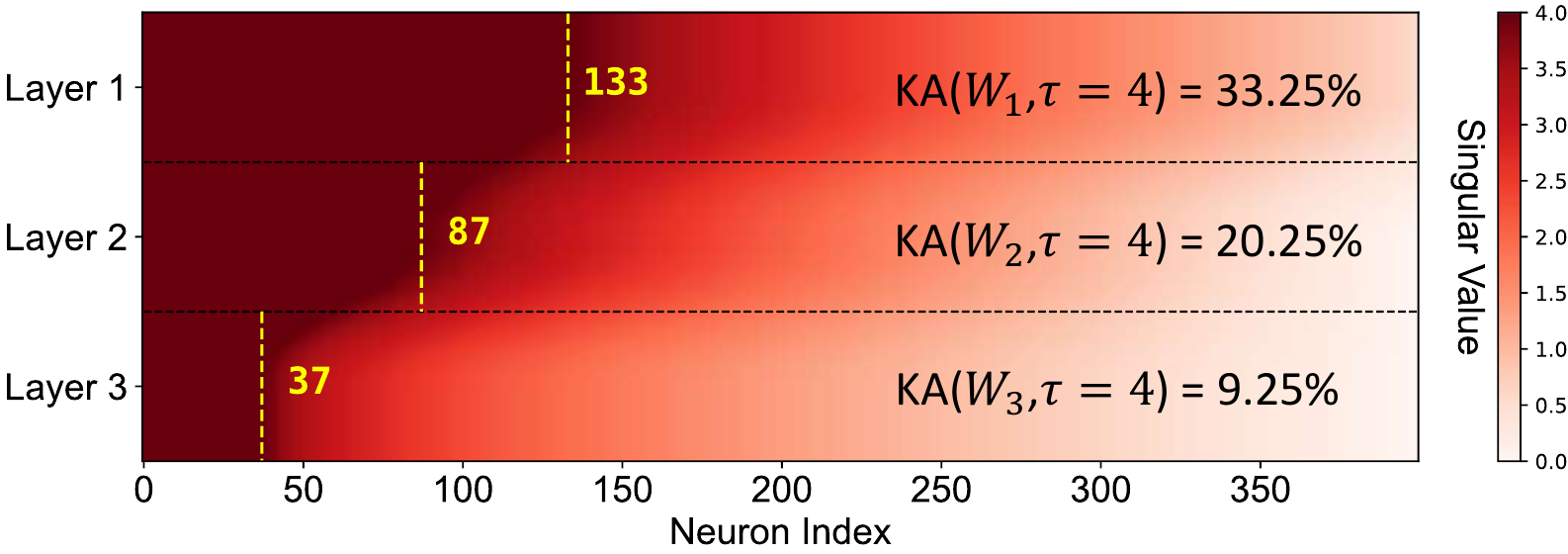}
        \subcaption{\footnotesize Singular value of all neurons in $1\mathbf{x}$ ensemble network.}
    \end{minipage}
    \begin{minipage}[t]{0.49\linewidth}
        \centering
        \includegraphics[width=\linewidth]{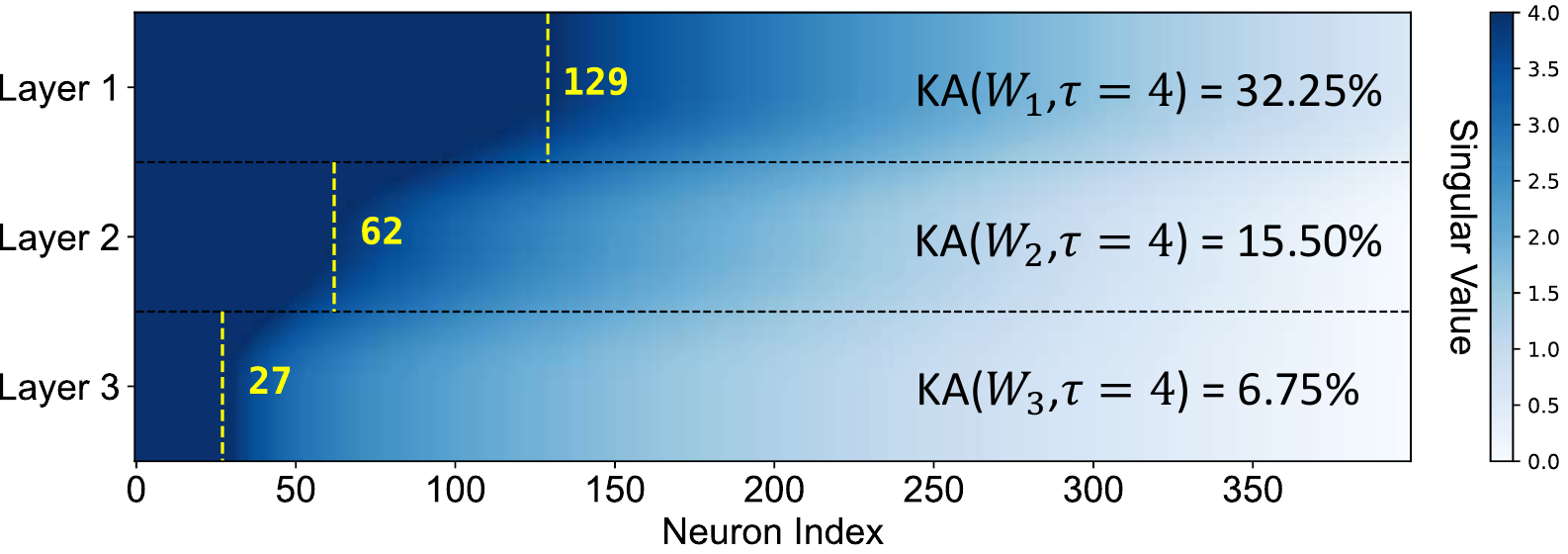}
        \subcaption{\footnotesize Singular value in the best sub-network of $6\mathbf{x}$ ensemble network.}
    \end{minipage}%
    \captionsetup{justification=raggedright}
    \caption{Singular value distributions of ensemble networks of different scales on the Criteo dataset, clipped in the range [0, 4]. The yellow dashed line indicates the index of the last neuron in the current layer whose singular value exceeds 4.}
    \label{Net_collapse_all}
\vspace{-1em}
\end{figure*}

\begin{tcolorbox}[colframe=black!75!white, colback=white!95!blue, arc=3mm, boxrule=0.5mm, width=\linewidth]
\textbf{Cause:} Knowledge abundance decreases not only with increasing network depth but also with the number of sub-networks. This phenomenon significantly undermines the diversity and informativeness of feature representations, thereby limiting the performance of the CTR prediction task.
\end{tcolorbox}

Based on the above cause finding, we propose a further hypothesis. As illustrated in Fig. \ref{moti} (c), the sub-networks within the ensemble architecture are trained in a \textit{parallel and independent manner}, relying solely on the 1-bit supervision signal derived from the ground-truth label $y$. Therefore, we attribute the aforementioned limitations (Findings 1 to 3) to two architectural design factors: (1) the lack of knowledge exchange among sub-networks reduces their KA; and (2) supervising multiple sub-networks with only a 1-bit signal provides overly limited supervision, which fails to effectively guide each sub-network to learn diverse and complementary discriminative features representation, especially in deeper layers. Consequently, knowledge transfer methods hold promise for mitigating dimensional collapse and enhancing the overall effectiveness of ensemble architectures.

Several related studies further support our perspective. Allen-Zhu \textit{et al.} \cite{dark_knowledge} point out that ensemble networks often contain \textit{dark knowledge}, i.e., implicit inter-class relational information implied in the output distribution of the ensemble network (a.k.a. soft labels), which is difficult to obtain under 1-bit one-hot label supervision. Their work further demonstrates that transferring such \textit{dark knowledge} via knowledge distillation significantly improves the performance of student networks, particularly in terms of feature diversity and generalization capability. In addition, Zhang \textit{et al.} \cite{DML} show that introducing mutual learning signals among sub-networks enhances the complementarity and diversity of their representations, thereby improving ensemble performance. These works underscore the importance of knowledge exchange among sub-networks from different perspectives.

Therefore, introducing knowledge distillation and mutual learning among sub-networks has the potential to promote KA of networks, which may help address the limitations identified in Findings 1 to 3.

\subsection{Enhancing Ensemble Network via Knowledge Distillation}
\label{section kd}
\begin{figure}[t]
\centering
    \begin{minipage}[t]{0.45\linewidth}
        \centering
        \includegraphics[width=\linewidth]{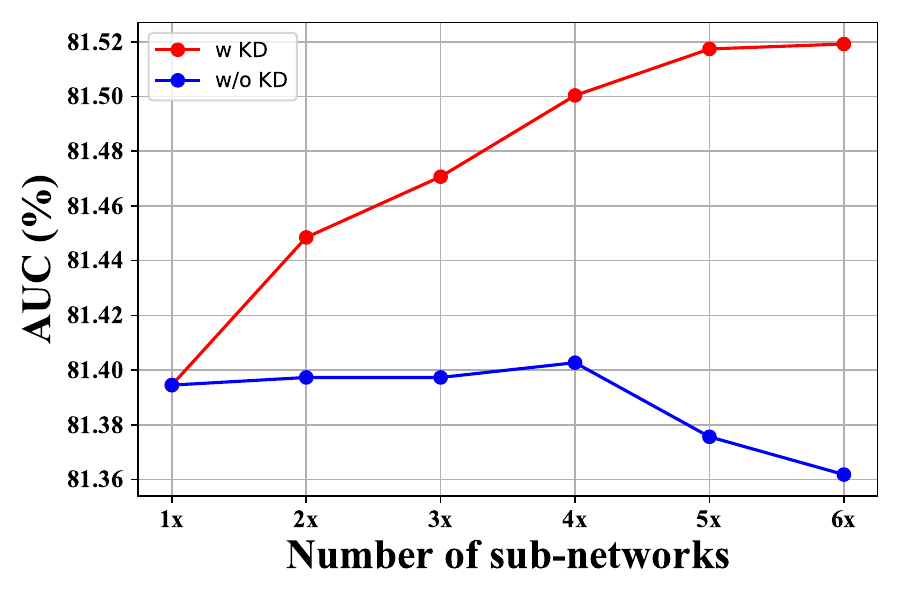}
        \subcaption{\footnotesize Ensemble prediction result $\hat{y}$}
    \end{minipage}
    \hspace{-0.5em} 
    \begin{minipage}[t]{0.45\linewidth}
        \centering
        \includegraphics[width=\linewidth]{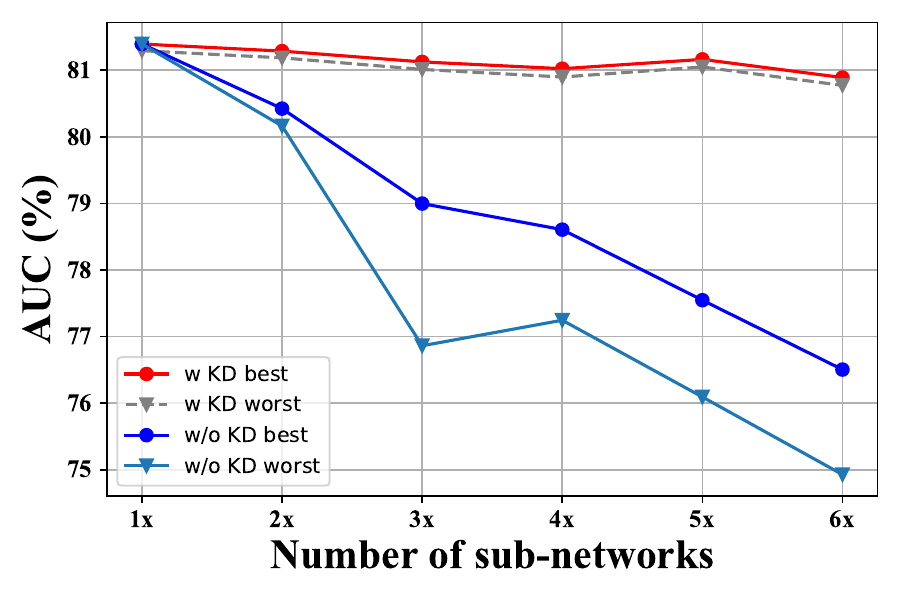}
        \subcaption{\footnotesize Best and worst sub-network}
    \end{minipage}%
    \captionsetup{justification=raggedright}
    \caption{The performance changes of the ensemble network enhanced by knowledge distillation on the Criteo dataset.}
    \label{KDs}
    \vspace{-1em}
\end{figure}

The core idea of knowledge distillation \cite{KD_survey1, KD_survey2} is to transfer knowledge from a complex and high-performing teacher model to a student model (i.e., teacher-to-student), thereby enhancing the latter's performance. However, designing and training an effective teacher model is a complex and time-consuming process. It requires significant computational resources and carefully tuned hyperparameters to ensure the teacher model provides a sufficient performance advantage when guiding student models. Consequently, we attempt to use the ensemble predictions results $\hat{y}$ as an abstract teacher to guide the learning process of each student (sub-network). The total loss $\mathcal{L}_{KD-CTR}$ is as follows:
\begin{equation}  
\begin{aligned}
\label{KD}
\mathcal{L}_{KD-CTR} &= \mathcal{L}_{CTR}\left(\hat{y}_{t}\right) + \sum_{k=1}^K \mathcal{L}_{KD}^k, \\
\hat{y}_{t} &= \texttt{Mean}(\hat{y}_{s,1}, \hat{y}_{s,2}, \dots, \hat{y}_{s,k}),\\
\mathcal{L}_{CTR}\left(\hat{y}_{t}\right)&=-\frac{1}{N} \sum_{i=1}^N\left(y_i \log \left(\hat{y}_{t,i}\right)+\left(1-y_i\right) \log \left(1-\hat{y}_{t,i}\right)\right), \\
\mathcal{L}_{KD}^k&=\lambda_k \cdot \mathcal{L}_{MSE}\left(\hat{y}_{t}, \hat{y}_{s,k}\right)=\frac{\lambda_k}{N} \sum_{i=1}^N\left(\hat{y}_{t,i} - \hat{y}_{s,k,i}\right)^2, 
\end{aligned}
\end{equation}
where $y$ denotes the true labels, $N$ denotes the batch size, $K$ is the number of sub-networks, $\lambda$ is a hyperparameter that balances the weight of the loss, $y_{t,i}$ represents the prediction result of the teacher model for the $i$-th sample, and $\hat{y}_{s,k,i}$ denotes the prediction result of the $k$-th student model for the $i$-th sample. From Equation \ref{KD}, it can be observed that due to the introduction of $\mathcal{L}_{KD}$, the different student networks (sub-networks) not only receive guidance from the true labels $y$ but also supervision signals from the teacher model $\hat{y}$, which is an abstract entity formed by collective decision-making of the student models. Therefore, this method of knowledge distillation provides additional supervision signals for student models. To empirically validate its effectiveness, we use the more stable mean operation as the abstract teacher. The specific experimental results are shown in Fig. \ref{KDs}. We have the following conclusions:

\begin{tcolorbox}[colframe=black!75!white, colback=white!95!blue, arc=3mm, boxrule=0.5mm, width=\linewidth]
Finding 4 (Solution). \textbf{KD is good medicine for Finding 1 \& 2}: As shown in Fig. \ref{KDs} (a), KD enables the ensemble model to follow scaling laws, improving performance with increasing parameters, addressing Finding 1. Additionally, Fig.~\ref{KDs} (b) indicates that it moderately reduces sub-network performance decline and variance, as noted in Finding 2.
\end{tcolorbox}

\begin{figure}[t]
\centering
    \begin{minipage}[t]{0.45\linewidth}
        \centering
        \includegraphics[width=\linewidth]{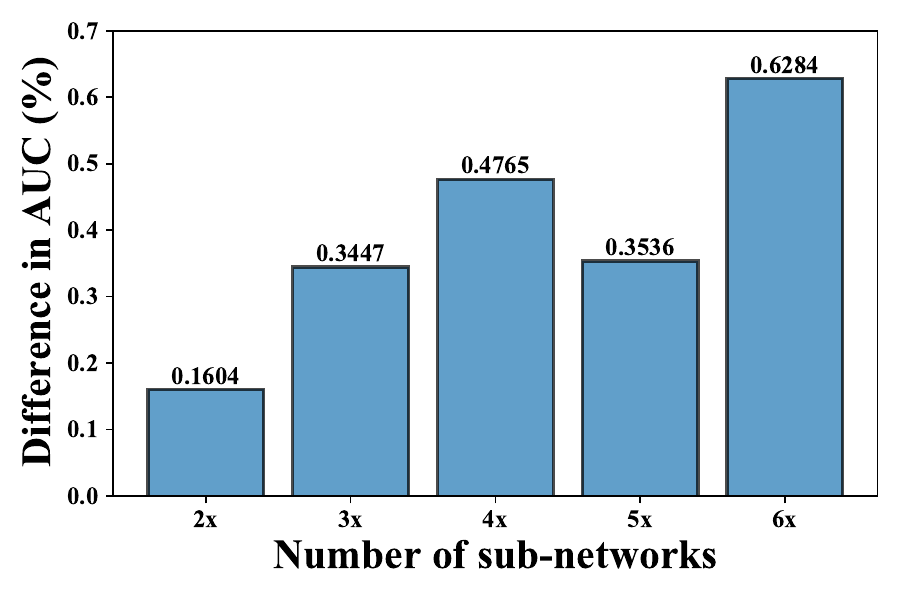}
        \subcaption{\footnotesize KD}
    \end{minipage}
    \hspace{-0.5em} 
    \begin{minipage}[t]{0.45\linewidth}
        \centering
        \includegraphics[width=\linewidth]{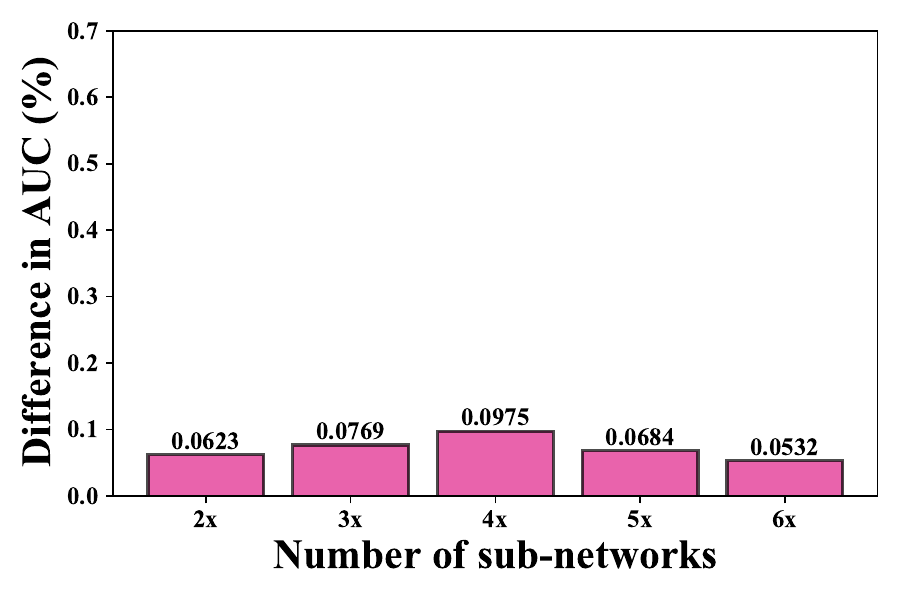}
        \subcaption{\footnotesize DML}
    \end{minipage}%
    \captionsetup{justification=raggedright}
    \caption{The performance differences between the best sub-network $\hat{y}_{s}$ and the ensemble prediction $\hat{y}$.}
    \label{Difference}
    \vspace{-1em}
\end{figure}

Although knowledge distillation via teacher-to-student effectively addresses the limitations mentioned in Findings 1 \& 2, as demonstrated in Fig. \ref{Difference} (a), it still struggles to resolve Finding 3, which concerns the substantial performance gap between the teacher and student models. Therefore, we further explore the impact of deep mutual learning via peer-to-peer on sub-network performance.

\subsection{Enhancing Ensemble Network via
Deep Mutual Learning}
\label{section dml}
The core idea of deep mutual learning \cite{DML} aims to facilitate knowledge transfer between peer-to-peer, where each student model serves as a teacher to others. Meanwhile, each sub-network directly receives supervision signals from true labels $y$. Unlike traditional knowledge distillation, which relies on a teacher model, this method allows for direct communication between sub-networks. The total loss $\mathcal{L}_{DML-CTR}$ is as follows:
\begin{equation}  
\begin{aligned}
\label{DML}
\mathcal{L}_{DML-CTR} &= \sum_{k=1}^K \left( \mathcal{L}_{CTR}\left(\hat{y}_{s,k}\right) + \mathcal{L}_{DML}^k \right), \\
\mathcal{L}_{DML}^k&= \sum_{l=1,l\neq k}^K \mu_{k,l} \cdot \mathcal{L}_{MSE}\left(\hat{y}_{s,l}, \hat{y}_{s,k}\right),
\end{aligned}
\end{equation}
where $\mu_{k,l}$ represents the intensity of knowledge that the $k$-th sub-network acquires from its multiple peers $l$\footnote{In Section \ref{section dml}, we follow the \cite{DML} and set $\mu = \frac{1}{K}$.}. From Equation \ref{DML}, it is evident that the introduction of $\mathcal{L}_{DML}^k$ enhances direct communication among student models, which helps to mitigate parallel and independent limitations between multiple sub-networks. To empirically assess its impact on ensemble network, we conduct experimental validation, the results of which are shown in Fig. \ref{DMLs} and \ref{Difference} (b). Our findings include the following:

\begin{figure}[t]
\centering
    \begin{minipage}[t]{0.45\linewidth}
        \centering
        \includegraphics[width=\linewidth]{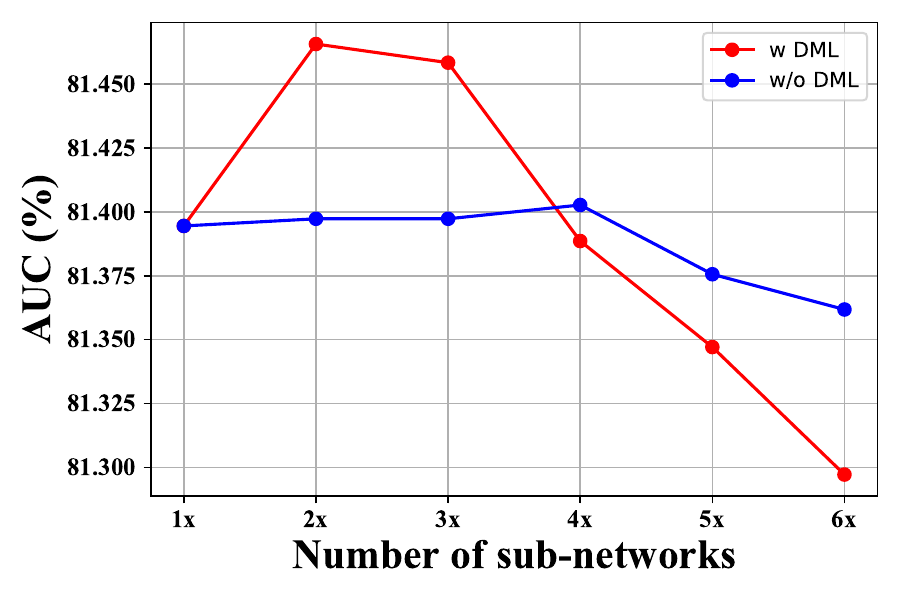}
        \subcaption{\footnotesize Ensemble prediction result $\hat{y}$}
    \end{minipage}
    \hspace{-0.5em} 
    \begin{minipage}[t]{0.45\linewidth}
        \centering
        \includegraphics[width=\linewidth]{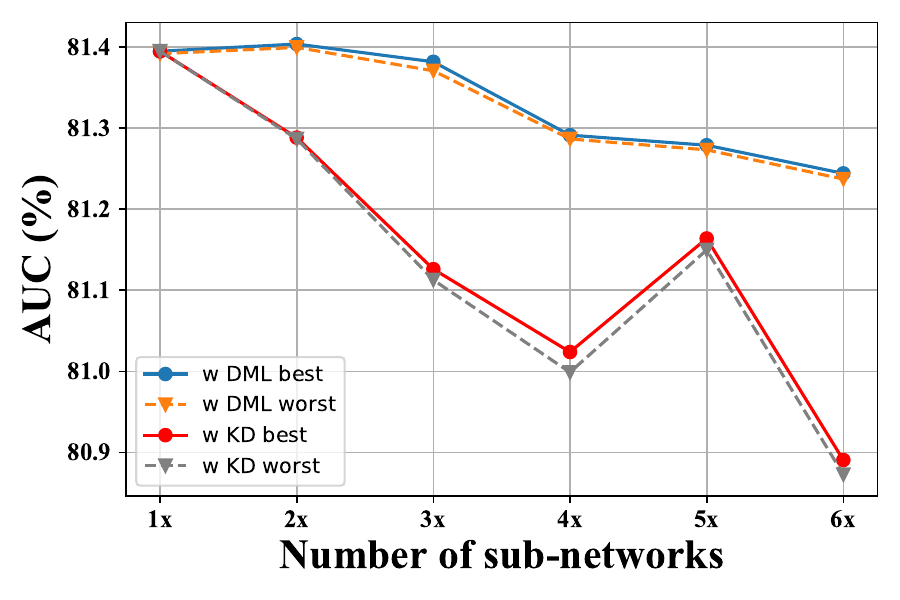}
        \subcaption{\footnotesize Best and worst sub-network}
    \end{minipage}%
    \captionsetup{justification=raggedright}
    \caption{The performance changes of ensemble network enhanced by deep mutual learning on the Criteo dataset.}
    \label{DMLs}
    \vspace{-1em}
\end{figure}

\begin{figure*}[t]
  \centering
  \includegraphics[width=1\textwidth]{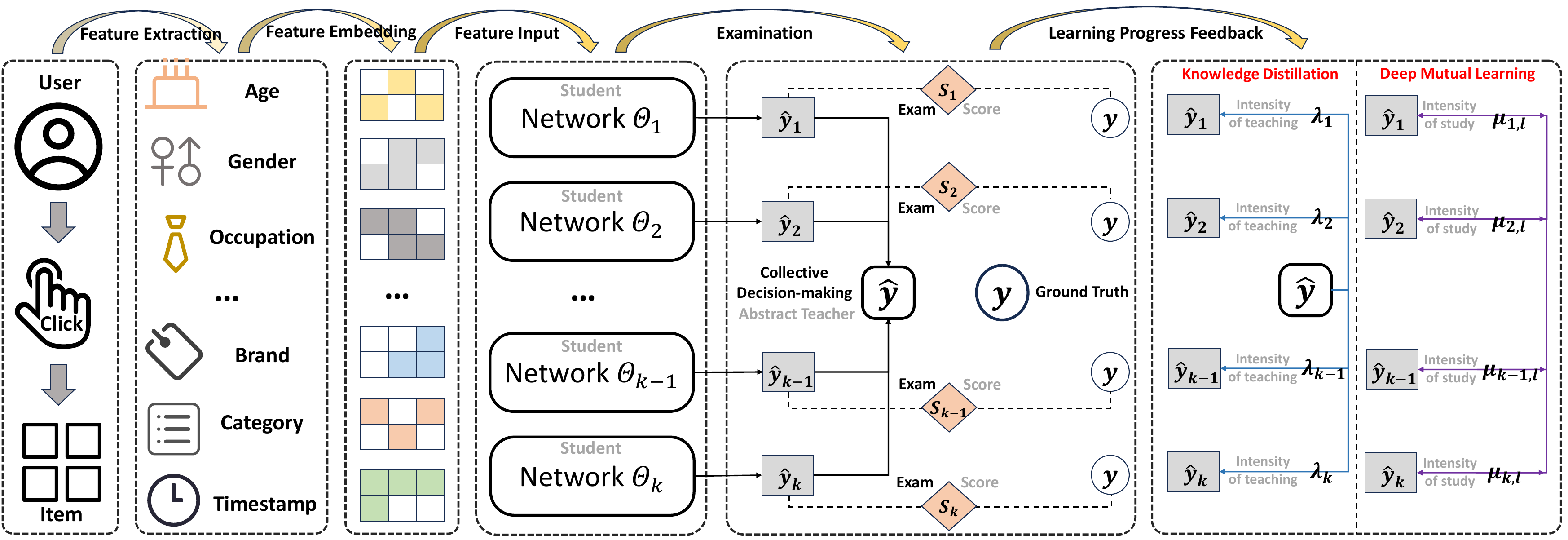}
  \caption{The workflow for the knowledge-driven ensemble framework (KDEF).}
  \label{KDEF}
  \vspace{-1em}
\end{figure*}

\begin{tcolorbox}[colframe=black!75!white, colback=white!95!blue, arc=3mm, boxrule=0.5mm, width=\linewidth]
Finding 5 (Solution). \textbf{DML is good medicine for Finding 2 \& 3}: Fig.~\ref{DMLs} shows that DML, unlike KD, does not address Finding 1’s limitation but better reduces sub-network performance degradation and variance (Finding 2). Meanwhile, DML also closes the performance gap between the ensemble network and sub-network, addressing Finding 3, as confirmed by Fig.~\ref{Difference} (b).
\end{tcolorbox}

To address our Findings 1 to 3 simultaneously, a simple idea is to combine KD with DML, where students are guided by both an abstract teacher and true labels, while also learning from each other. However, this simple idea introduces an additional limitation: 

\begin{tcolorbox}[colframe=black!75!white, colback=white!95!blue, arc=3mm, boxrule=0.5mm, width=\linewidth]
Finding 6 (limitation). \textbf{Manual parameter tuning is impractical.} Equations (\ref{KD}) and (\ref{DML}) both require the loss balancing hyperparameters $\lambda$ and $\mu$. The search spaces for these hyperparameters are $O(K)$ and $O(K(K-1))$ respectively, making manual tuning of these hyperparameters impractical. Therefore, a new adaptive loss balancing mechanism is needed to facilitate tailored teaching from teacher-to-student and selective learning in peer-to-peer. 
\vspace{-0.5em}
\end{tcolorbox}

\subsection{Examination Mechanism}
To address Finding 6, we propose an examination mechanism that utilizes the absolute difference between the predictions of the sub-networks and the true labels as a quantitative measure of model learning progress. The formulation is as follows:
\begin{equation}  
\begin{aligned}
S_k&= \frac{1}{N} \sum_{i=1}^N \left(1 - \left\lVert y_i - \hat{y}_{s,k,i} \right\rVert \right), 
\end{aligned}
\end{equation}
where $S_k$ represents the examination score of the $k$-th sub-network, a smaller $S_k$ indicates better learning progress for that sub-network, and vice versa. Intuitively, when a teacher identifies a student with a low score, the mechanism enhances the instructional intensity directed toward that student. Simultaneously, during peer-to-peer learning, students should preferentially learn from their more capable peers while reducing the influence of lower-performing ones to prevent knowledge conflicts \cite{KDCL}. The introduction of the examination mechanism into $\lambda$ and $\mu$ is shown as follows:
\begin{equation}  
\begin{aligned}
\label{loss_weight}
\lambda_k &= \text{Softmin}(S_k) = \frac{\exp \left(-S_k\right)}{\sum_{i=1}^K \exp \left(-S_i\right)}, \\
\mu_{k,l} &= \text{Softmax}(S_l - S_k) = \frac{\exp \left(S_l - S_k\right)}{\sum_{l=1, l\neq k}^K \exp \left(S_l - S_k\right)},
\end{aligned}
\end{equation}

\subsection{Knowledge-Driven Ensemble Framework}
\label{embedding}

\begin{figure*}[t]
\centering
    \begin{minipage}[t]{0.33\linewidth}
        \centering
        \includegraphics[width=\linewidth]{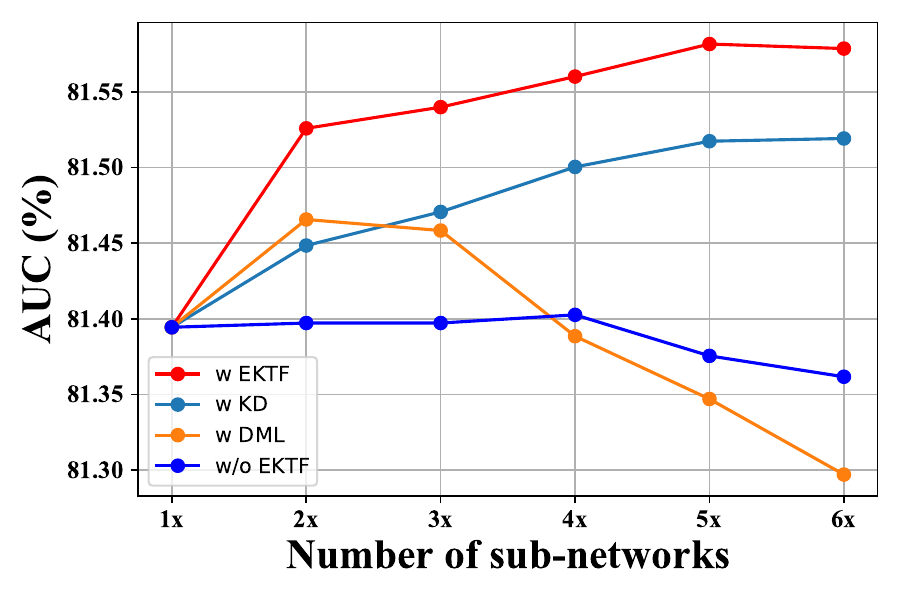}
        \subcaption{\footnotesize Ensemble prediction result $\hat{y}$}
    \end{minipage}
    \hspace{-0.5em} 
    \begin{minipage}[t]{0.33\linewidth}
        \centering
        \includegraphics[width=\linewidth]{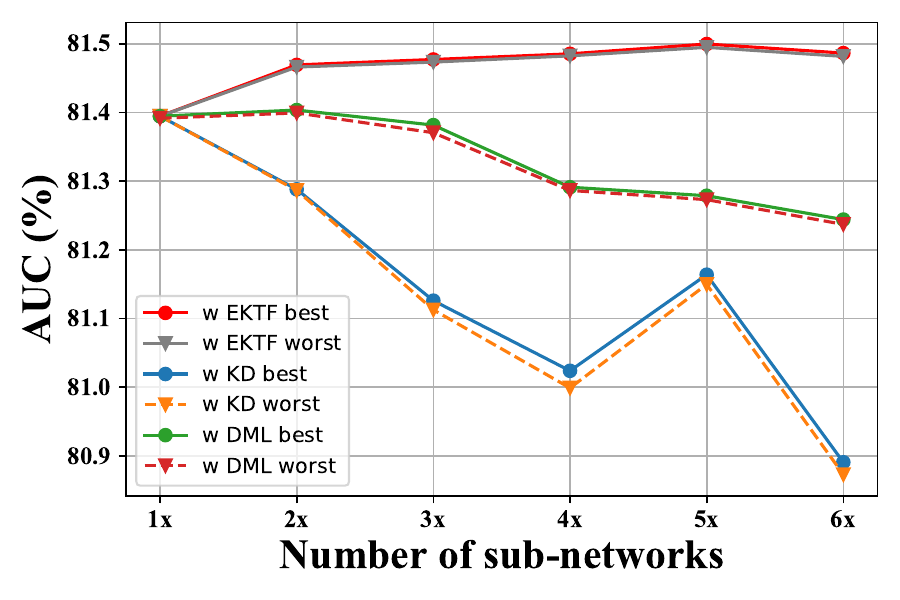}
        \subcaption{\footnotesize Best and worst sub-network}
    \end{minipage}
    \hspace{-0.5em} 
    \begin{minipage}[t]{0.33\linewidth}
        \centering
        \includegraphics[width=\linewidth]{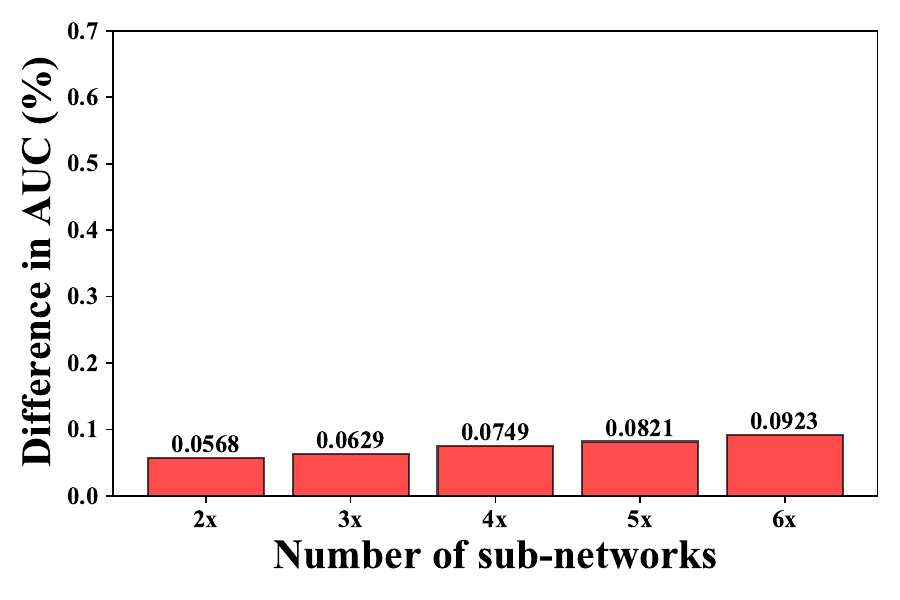}
        \subcaption{\footnotesize \footnotesize The performance differences between $\hat{y}$ and best sub-network $\hat{y}_{s}$}
    \end{minipage}%
    \captionsetup{justification=raggedright}
    \caption{Comparison of KDEF with KD and DML in addressing Finding 1 to 3 on the Criteo dataset.}
    \label{KDEF_solution}
\end{figure*}

\begin{figure}[t]
\centering
    \begin{minipage}[t]{1\linewidth}
        \centering
        \includegraphics[width=\linewidth]{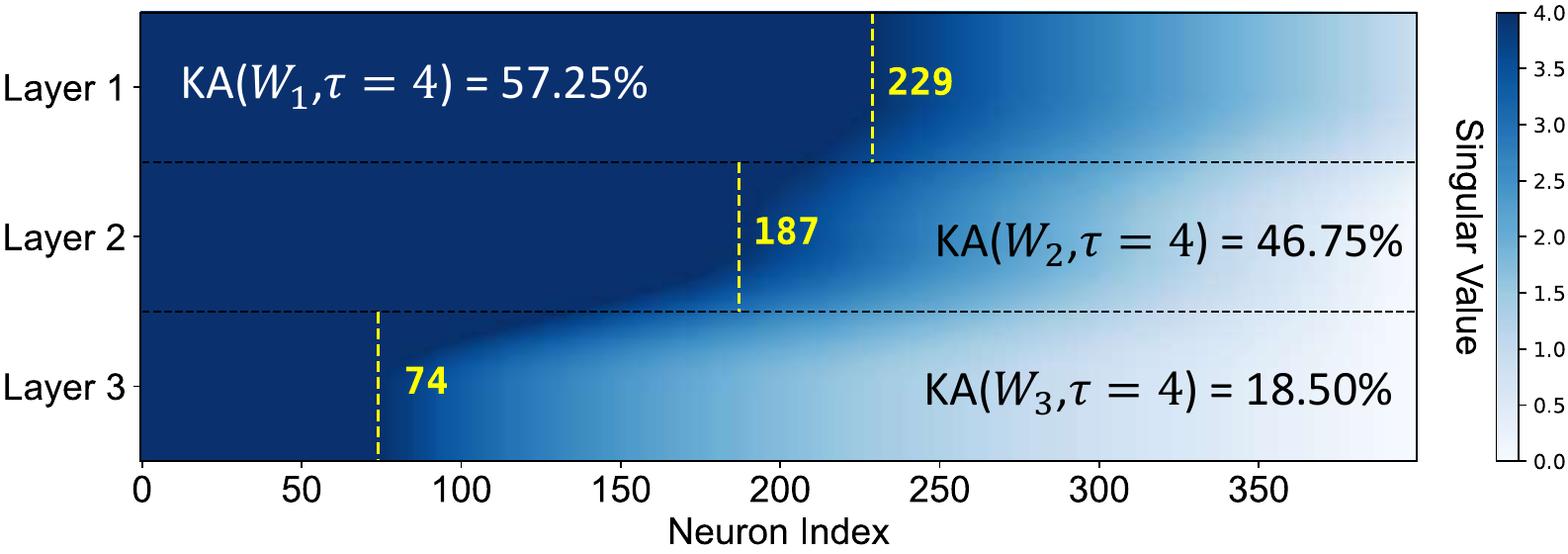}
    \end{minipage}
    \captionsetup{justification=raggedright}
    \caption{Singular value in the best sub-network of KDEF$_{10\mathbf{x}}$ on the Criteo dataset.}
    \label{KDEF_SVD}
\end{figure}

By integrating the above methods, we propose a knowledge-driven ensemble framework (KDEF) that can be expanded to include an infinite number of student networks. The specific architecture is illustrated in Fig. \ref{KDEF}. Initially, we extract the following three types of data and convert them into multi-field categorical features, which are then subjected to one-hot encoding:
\begin{itemize}[leftmargin=*]
\item  \emph{User profiles} ($x_U$): age, gender, occupation, etc.
\item \emph{Item attributes} ($x_I$): brand, price, category, etc.
\item \emph{Context} ($x_C$): timestamp, device, position, etc.
\end{itemize}
We can define a CTR input sample in the tuple data format: $X = \{x_U, x_I, x_C\}$. Further, most CTR prediction models \cite{autoint, CETN, adagin} utilize an embedding layer to transform them into low-dimensional dense vectors: $\mathbf{e}_i=\textit{E}_i x_i$, where $\textit{E}_i \in \mathbb{R}^{d \times s_i}$ and $s_i$ separately indicate the embedding matrix and the vocabulary size for the $i$-th field, $d$ represents the embedding dimension. After that, we concatenate the individual features to get the input $\mathbf{h}=\left[\mathbf{e}_1, \mathbf{e}_2, \cdots, \mathbf{e}_f\right]$ to the student networks. Upon receiving the prediction results $\hat{y}_{s,k}=\texttt{Network}_{k}(\mathbf{h})$ from each student network and the true labels, we integrate the ideas of KD and DML, incorporating the examination mechanism to derive our final loss function:
\begin{equation}  
\begin{aligned}
\label{loss_KDEF}
\mathcal{L}_{KDEF} &= \frac{1}{K}\sum_{k=1}^K \mathcal{L}_{CTR}\left(\hat{y}_{s,k}\right) + \sum_{k=1}^K \mathcal{L}_{KD}^k + \frac{1}{K}\sum_{k=1}^K \mathcal{L}_{DML}^k.
\end{aligned}
\end{equation}

The complete training processes of KDEF are shown in Algorithm \ref{training_process}. To validate whether our proposed KDEF simultaneously addresses the limitations in Findings 1 to 3 above, we conduct further experiments. Fig. \ref{KDEF_solution} (a) demonstrates that the ensemble prediction result $\hat{y}$ of KDEF consistently outperforms both the single KD and DML, thereby resolving Finding 1. Fig. \ref{KDEF_solution} (b) indicates that the performance variations of KDEF's sub-network performance vary more smoothly and with less variance, thus solving Finding 2. Lastly, Fig. \ref{KDEF_solution} (c) shows that the prediction result $\hat{y}$ differs from the best student network's prediction $\hat{y}_s$ by less than 0.1\%, thus resolving Finding 3. Fig. \ref{KDEF_SVD} demonstrates that our KDEF significantly mitigates the dimensional collapse phenomenon. For example, in the third layer of the $10\mathbf{x}$ network, the knowledge abundance reaches 18.5\%, whereas the $1\mathbf{x}$ network, as shown in Fig. \ref{Net_collapse_all} (a), exhibits only 9.25\%. Meanwhile, we further average $\mathcal{L}_{CTR}$ and $\mathcal{L}_{DML}$ rather than simply summing them. This approach ensures that the training is predominantly guided by the stable teacher model and helps reduce the instability caused by fluctuations in early loss. Besides, it is worth noting that if we can reduce the performance gap between the sub-networks $\hat{y}_s$ and the ensemble prediction $\hat{y}$, we can achieve a more flexible ensemble model. When higher performance is required, we can employ the ensemble model; conversely, when lower model complexity is needed, we can utilize the best-performing student model.

\begin{algorithm}[t]
\caption{The training process of KDEF\label{training_process}}
  \SetAlgoLined 
  \KwIn{input samples $X \in N$;}
  \KwOut{model parameters $\Theta$; }
  Initialize parameters $\Theta$\;
  \While{KDEF has not reached the early stopping patience threshold}{
   \For{$X \in N$}{
       Calculate feature embeddings $\mathbf{h}$ according to Section \ref{embedding}\;
        Calculate all sub-network predictions to get \{$\hat{y}_{s,1}, \hat{y}_{s,2}, \dots, \hat{y}_{s,k}$\} \;
        Calculate ensemble prediction to get $\hat{y}_t$ according to Eq. (\ref{ensemble_prediction})\;
    }
    Calculate loss weights $\lambda_k$ and $\mu_{k,l}$ according to Eq. (\ref{loss_weight})\;
    Calculate total loss $\mathcal{L}_{KDEF}$ according to Eq. (\ref{loss_KDEF})\;
   Update parameters $\Theta$ by descending the gradients $\nabla_{\Theta} \mathcal{L}_{KDEF}$;
    }
    \Return model parameters $\Theta$;
\end{algorithm}

\section{Experiments}
\label{Experiments}
In this section, we conduct comprehensive experiments on six CTR prediction datasets to validate the effectiveness and compatibility of our proposed KDEF. We aim to address the following research questions (RQs):
\begin{itemize}[leftmargin=*]
\item \textbf{RQ1} Does KDEF enable the CTR model to achieve an ensemble scaling law?
\item \textbf{RQ2} If heterogeneous networks are used as student networks in KDEF, how do they perform? Do they perform well on large-scale and highly sparse datasets?
\item \textbf{RQ3} Does KDEF outperform other knowledge transfer methods?
\item \textbf{RQ4} How do the various components of the KDEF affect performance?
\item \textbf{RQ5} How are KDEF compatibility, efficiency, and flexibility?

\end{itemize}

\subsection{Experiment Setup}
\subsubsection{\textbf{Datasets.}} We evaluate KDEF on five CTR prediction datasets: Criteo\footnote{\url{https://www.kaggle.com/c/criteo-display-ad-challenge}} \cite{openbenchmark}, ML-1M\footnote{\url{https://grouplens.org/datasets/movielens}} \cite{autoint}, KDD12\footnote{\url{https://www.kaggle.com/c/kddcup2012-track2}} \cite{autoint}, iPinYou\footnote{\url{https://contest.ipinyou.com/}} \cite{pnn2}, and KKBox\footnote{\url{https://www.kkbox.com/intl}} \cite{Bars}. Table \ref{dataset}  provides detailed information about these datasets. A more detailed description of these datasets can be found in the given references and links.

\begin{table}[t]
\tiny
\renewcommand\arraystretch{1}
\centering
\caption{Dataset statistics}
\label{dataset}
\resizebox{0.9\linewidth}{!}{
\begin{tabular}{cccc} 
\toprule 
\textbf{Dataset} & \textbf{\#Instances} & \textbf{\#Fields} & \textbf{\#Features} \\
\midrule 
\textbf{Criteo} & 45,840,617 & 39  & 910,747 \\
\textbf{ML-1M} & 739,012  & 7  & 9,751 \\
\textbf{KDD12} & 141,371,038  & 13  & 4,668,096 \\
\textbf{iPinYou} & 19,495,974 & 16  & 665,765\\
\textbf{KKBox} & 7,377,418 & 13  & 91,756\\
\bottomrule
\end{tabular}}
\end{table}

\subsubsection{\textbf{Data Preprocessing.}} We follow the approach outlined in \cite{openbenchmark}. For the Criteo and KDD12 dataset, we discretize the numerical feature fields by rounding down each numeric value $x$ to $\lfloor \log^2(x) \rfloor$ if $x > 2$, and $x = 1$ otherwise. We set a threshold to replace infrequent categorical features with a default "OOV" token. We set the threshold to 10 for Criteo, KKBox, and KDD12, 2 for iPinYou, and 1 for the small dataset ML-1M. More specific data processing procedures and results can be found in our open-source run logs\footnote{\url{https://github.com/salmon1802/KDEF/tree/main/checkpoints/} \label{footnote:checkpoint}} and configuration files, which we do not elaborate on here.

\subsubsection{\textbf{Evaluation Metrics.}} To compare the performance, we utilize two commonly used metrics in CTR models: \textbf{Logloss}, \textbf{AUC} \cite{autoint, GDCN, ComboFashion}. AUC stands for Area Under the ROC Curve, which measures the probability that a positive instance will be ranked higher than a randomly chosen negative one. Logloss is the result of the calculation of $\mathcal{L}_{CTR}$. A lower Logloss suggests a better capacity for fitting the data.

\subsubsection{\textbf{Baselines.}}
\label{section_baseline}
We compared $\textbf{KDEF}_{\text{MLP}}$ and $\textbf{KDEF}_{\text{Lastest}}$ with some SOTA models (\textbf{*} denotes integrating the original model with DNN networks). $\textbf{KDEF}_{\text{MLP}}$ refers to using homogeneous MLPs as the sub-networks (students), while $\textbf{KDEF}_{\text{Latest}}$ refers to using the three latest CTR models as sub-networks (in this paper, PNN, FINAL, and ECN are selected). Further, we select several high-performance CTR representative baselines, such as PNN \cite{pnn1} and Wide \& Deep \cite{widedeep} (2016); DeepFM \cite{deepfm} and DCNv1 \cite{dcn} (2017); xDeepFM (2018) \cite{xdeepfm}; AutoInt* (2019) \cite{autoint}; AFN* (2020) \cite{AFN}; DCNv2 \cite{dcnv2} and EDCN \cite{EDCN}, MaskNet \cite{masknet} (2021); CL4CTR \cite{CL4CTR}, EulerNet \cite{EulerNet}, FinalMLP \cite{finalmlp}, FINAL \cite{FINAL} (2023), RFM \cite{RFM}, and ECN \cite{FCN} (2024). For knowledge transfer methods, we select four representative baselines: Online KD \cite{KD} (2015), KDCL \cite{KDCL} (2020), ECKD \cite{KD4CTR} (2020), and DAGFM \cite{DAGFM} (2023).

\subsubsection{\textbf{Implementation Details.}} We implement all models using PyTorch \cite{PYTORCH} and refer to existing works \cite{openbenchmark, FuxiCTR}. We employ the Adam optimizer \cite{adam} to optimize all models, with a default learning rate set to 0.001. For the sake of fair comparison, we set the embedding dimension to 128 for KKBox and 16 for the other datasets \cite{openbenchmark, Bars}. The batch size is set to 4,096 on the ML-1M, iPinYou datasets, and 10,000 on the other datasets. The default DNN architecture is [400, 400, 400]. The Dropout rate is determined via grid search over the set \{0.1, 0.2, 0.3\}. During training, we employ a Reduce-LR-on-Plateau scheduler that reduces the learning rate by a factor of 10 when performance stops improving in any given epoch. To prevent overfitting, we apply early stopping on the validation set with a patience of 2 epochs \cite{finalmlp, openbenchmark}. The hyperparameters of the baseline model are configured and fine-tuned based on the \textit{optimal values} provided in  \cite{FuxiCTR,openbenchmark} and their original paper. Further details on model hyperparameters and dataset configurations are available in our straightforward and accessible running logs\footref{footnote:checkpoint} and are not reiterated here.

\subsection{From Collapse to Stability: Scaling Up CTR Models (RQ1)}
\begin{table*}[t]
\renewcommand\arraystretch{1.2}
\centering
\caption{AUC performance comparison of scaled CTR models (higher is better). Underline and bold indicate the best performance achieved by mean fusion and KDEF, respectively. Mean$_{b}$ and KDEF${_b}$ denote the best-performing sub-networks in mean fusion and KDEF, respectively. Typically, CTR researchers consider an improvement of \textit{0.001 (0.1\%)} in AUC to be statistically significant \cite{dcn,EDCN,CL4CTR,openbenchmark}. } 
\label{scaling_up}
\resizebox{\linewidth}{!}{
\begin{tabular}{cc|ccccccc|ccccccc}
\hline \Xhline{1px}
\multicolumn{2}{c|}{\multirow{2}{*}{Model}} &
  \multicolumn{7}{c|}{Criteo} &
  \multicolumn{7}{c}{KKBox} \\ \cline{3-16} 
\multicolumn{2}{c|}{} &
  base &
  $2\mathbf{x}$ &
  $3\mathbf{x}$ &
  $4\mathbf{x}$ &
  $5\mathbf{x}$ &
  $6\mathbf{x}$ &
  $10\mathbf{x}$ &
  base &
  $2\mathbf{x}$ &
  $3\mathbf{x}$ &
  $4\mathbf{x}$ &
  $5\mathbf{x}$ &
  $6\mathbf{x}$ &
  $10\mathbf{x}$  \\ \hline
\multicolumn{1}{c|}{\multirow{4}{*}{MLP}} &
  Mean &
  \multirow{2}{*}{81.39} &
  81.39 &
  81.39 &
  {\ul 81.40} &
  81.37 &
  81.36 &
  81.25 &
  \multirow{2}{*}{85.01} &
  84.97 &
  84.89 &
  {\ul 85.11} &
  84.94 &
  83.90 &
  84.89 \\
\multicolumn{1}{c|}{} &
  \textbf{KDEF} &
   &
  81.52 &
  81.54 &
  81.56 &
  81.58 &
  81.58 &
  \textbf{81.60} &
   &
  85.28 &
  85.51 &
  85.49 &
  85.47 &
  85.02 &
  \textbf{85.66} \\ \cline{2-16} 
\multicolumn{1}{c|}{} &
  Mean$_{b}$ &
  \multirow{2}{*}{{\ul 81.39}} &
  80.42 &
  78.99 &
  78.60 &
  77.54 &
  76.50 &
  74.49 &
  \multirow{2}{*}{{\ul 85.01}} &
  82.82 &
  81.07 &
  79.37 &
  77.11 &
  76.79 &
  74.89 \\
\multicolumn{1}{c|}{} &
  \textbf{KDEF}$_{b}$ &
   &
  81.46 &
  81.47 &
  81.48 &
  81.49 &
  81.48 &
  \textbf{81.50} &
   &
  85.17 &
  85.35 &
  85.30 &
  85.27 &
  84.93 &
  \textbf{85.39} \\ \hline
\multicolumn{1}{c|}{\multirow{4}{*}{FINAL}} &
  Mean &
  \multirow{2}{*}{{\ul 81.44}} &
  81.41 &
  81.38 &
  81.33 &
  81.37 &
  81.32 &
  81.23 &
  \multirow{2}{*}{85.13} &
  85.18 &
  {\ul 85.23} &
  85.15 &
  85.11 &
  84.82 &
  84.85 \\
\multicolumn{1}{c|}{} &
  \textbf{KDEF} &
   &
  81.54 &
  81.57 &
  81.59 &
  81.58 &
  81.61 &
  \textbf{81.62} &
   &
  85.18 &
  85.12 &
  85.26 &
  \textbf{85.68} &
  85.62 &
  85.48 \\ \cline{2-16} 
\multicolumn{1}{c|}{} &
  Mean$_{b}$ &
  \multirow{2}{*}{{\ul 81.44}} &
  79.32 &
  79.15 &
  78.47 &
  78.43 &
  78.08 &
  77.28 &
  \multirow{2}{*}{{\ul 85.13}} &
  83.94 &
  82.57 &
  79.79 &
  79.03 &
  76.96 &
  73.83 \\
\multicolumn{1}{c|}{} &
  \textbf{KDEF}$_{b}$ &
   &
  81.50 &
  81.52 &
  81.51 &
  81.52 &
  81.51 &
  \textbf{81.53} &
   &
  85.13 &
  85.05 &
  85.16 &
  \textbf{85.41} &
  85.38 &
  85.28 \\ \hline
\multicolumn{1}{c|}{\multirow{4}{*}{ECN}} &
  Mean &
  \multirow{2}{*}{{\ul 81.55}} &
  81.43 &
  81.43 &
  81.38 &
  81.32 &
  81.35 &
  81.33 &
  \multirow{2}{*}{{\ul 85.40}} &
  85.30 &
  85.05 &
  85.27 &
  85.30 &
  85.32 &
  85.25 \\
\multicolumn{1}{c|}{} &
  \textbf{KDEF} &
   &
  81.57 &
  81.59 &
  81.60 &
  81.60 &
  \textbf{81.61} &
  \textbf{81.61} &
   &
  85.51 &
  85.61 &
  85.57 &
  85.61 &
  \textbf{85.80} &
  85.71 \\ \cline{2-16} 
\multicolumn{1}{c|}{} &
  Mean$_{b}$ &
  \multirow{2}{*}{{\ul 81.55}} &
  80.83 &
  80.37 &
  79.69 &
  78.98 &
  78.77 &
  77.64 &
  \multirow{2}{*}{{\ul 85.40}} &
  84.73 &
  84.08 &
  83.64 &
  82.97 &
  82.88 &
  81.23 \\
\multicolumn{1}{c|}{} &
  \textbf{KDEF}$_{b}$ &
   &
  81.55 &
  81.56 &
  \textbf{81.57} &
  \textbf{81.57} &
  \textbf{81.57} &
  \textbf{81.57} &
   &
  85.48 &
  85.56 &
  85.53 &
  85.55 &
  \textbf{85.60} &
  85.57 \\ \hline
\multicolumn{1}{c|}{\multirow{4}{*}{PNN}} &
  Mean &
  \multirow{2}{*}{{\ul 81.42}} &
  81.38 &
  81.37 &
  81.36 &
  81.32 &
  81.29 &
  81.28 &
  \multirow{2}{*}{85.07} &
  85.15 &
  {\ul 85.22} &
  85.17 &
  85.16 &
  85.14 &
  84.94 \\
\multicolumn{1}{c|}{} &
  \textbf{KDEF} &
   &
  81.49 &
  81.51 &
  81.54 &
  81.54 &
  81.54 &
  \textbf{81.55} &
   &
  85.22 &
  85.25 &
  85.33 &
  \textbf{85.45} &
  85.35 &
  \textbf{85.45} \\ \cline{2-16} 
\multicolumn{1}{c|}{} &
  Mean$_{b}$ &
  \multirow{2}{*}{{\ul 81.42}} &
  80.17 &
  79.49 &
  78.64 &
  77.64 &
  77.38 &
  74.71 &
  \multirow{2}{*}{85.07} &
  83.14 &
  {\ul 80.17} &
  79.83 &
  78.24 &
  76.31 &
  73.38 \\
\multicolumn{1}{c|}{} &
  \textbf{KDEF}$_{b}$ &
   &
  81.44 &
  81.45 &
  \textbf{81.47} &
  81.44 &
  \textbf{81.47} &
  \textbf{81.47} &
   &
  85.18 &
  85.20 &
  85.26 &
  \textbf{85.34} &
  85.25 &
  85.33 \\ \hline
  \Xhline{1px}
\end{tabular}}
\vspace{-1em}
\label{scale_law}
\end{table*}

To verify whether KDEF successfully enables the ensemble network to exhibit a scaling law, we conduct ensemble scaling on several representative baseline models using the Criteo and KKBox datasets. The experimental results are presented in Fig. \ref{scale_law}. For ensemble networks using only mean fusion, we observe a sharp degradation in both ensemble performance and individual sub-network performance as the number of sub-networks increases across all models. For example, the FINAL model achieves an AUC of 81.44 when trained independently on the Criteo dataset, but its ensemble AUC drops to 81.23 after being scaled to $10\mathbf{x}$, with the best-performing sub-network reaching only 77.28, indicating a considerable performance gap. This phenomenon is consistent with our observations in Section \ref{Collapse_Phenomenon}, where increasing the number of sub-networks tends to drive the model towards low-rank solutions, thereby impairing overall performance. 

In contrast, KDEF effectively mitigates this collapse effect and enables stable performance improvements as the number of sub-networks increases. For example, the FINAL model achieves an AUC of 81.62 when scaled to $10\mathbf{x}$ on the Criteo dataset, while PNN reaches 85.45 on the KKBox dataset under the same scaling. Both results surpass the performance of mean fusion by more than 0.001. These results demonstrate that KDEF enhances the scalability of CTR models and provides a simple yet effective solution to the issue of dimensional collapse in large ensemble networks. 

Additionally, we observe that on the KKBox dataset, models do not always achieve the best performance at the $10\mathbf{x}$ scaling level, whereas the opposite trend is evident on the Criteo dataset. We attribute this to the fact that KKBox is a relatively small-scale dataset, and models such as FINAL, ECN, and PNN possess stronger feature interaction capabilities compared to MLP \cite{openbenchmark}. As a result, even with KDEF, these models are more prone to overfitting under limited data. Meanwhile, prior studies \cite{GPT3.0} suggest that the validity of scaling laws often depends on the alignment between model capacity and the amount of training data. When the dataset is insufficient to support larger models, performance may degrade rather than improve.

\subsection{Overall Performance}
\subsubsection{\textbf{Performance Comparison with Different Deep CTR Models (RQ2).}}

\begin{table*}[t]
\renewcommand\arraystretch{1.2}
\centering
\caption{Performance comparison of different deep CTR models. Meanwhile, we conduct a two-tailed T-test to assess the statistical significance between our models and the best baseline ($\star$: $p <$  1e-3). Typically, CTR researchers consider an improvement of \textit{0.001 (0.1\%)} in Logloss and AUC to be statistically significant \cite{dcn,EDCN,CL4CTR,openbenchmark}. } 
\label{baselines}
\resizebox{\linewidth}{!}{
\begin{tabular}{ccccccccccc}
\Xhline{1px}
  \multicolumn{1}{c|}{} &
  \multicolumn{2}{c|}{\textbf{Criteo}} &
  \multicolumn{2}{c|}{\textbf{ML-1M}} &
  \multicolumn{2}{c|}{\textbf{KDD12}} &
  \multicolumn{2}{c|}{\textbf{iPinYou}} &
  \multicolumn{2}{c}{\textbf{KKBox}}\\ \cline{2-11} 
  \multicolumn{1}{c|}{\multirow{-2}{*}{\textbf{Models}}} &
  \multicolumn{1}{c}{Logloss$\downarrow$} &
  \multicolumn{1}{c|}{AUC(\%)$\uparrow$} & 
  \multicolumn{1}{c}{Logloss$\downarrow$} &
  \multicolumn{1}{c|}{AUC(\%)$\uparrow$} &
  \multicolumn{1}{c}{Logloss$\downarrow$} &
  \multicolumn{1}{c|}{AUC(\%)$\uparrow$} &
  \multicolumn{1}{c}{Logloss$\downarrow$} &
  \multicolumn{1}{c|}{AUC(\%)$\uparrow$} &
  \multicolumn{1}{c}{Logloss$\downarrow$} &
  \multicolumn{1}{c}{AUC(\%)$\uparrow$} \\
  \hline
  \multicolumn{1}{c|}{PNN \cite{pnn1}} &
  0.4378 &
  \multicolumn{1}{c|}{81.42} &
  0.3070 &
  \multicolumn{1}{c|}{90.42} &
  0.1504 &
  \multicolumn{1}{c|}{80.47} &
  0.005544 &
  \multicolumn{1}{c|}{78.13} &
  0.4801 &
  \multicolumn{1}{c}{85.07} \\
  \multicolumn{1}{c|}{Wide \& Deep \cite{widedeep}} &
  0.4376 &
  \multicolumn{1}{c|}{81.42} &
  0.3056 &
  \multicolumn{1}{c|}{90.45} &
  0.1504 &
  \multicolumn{1}{c|}{80.48} &
  0.005542 &
  \multicolumn{1}{c|}{78.09} &
  0.4852 &
  \multicolumn{1}{c}{85.04} \\
  \multicolumn{1}{c|}{DeepFM \cite{deepfm}} &
  0.4375 &
  \multicolumn{1}{c|}{81.43} &
  0.3073 &
  \multicolumn{1}{c|}{90.51} &
  0.1501 &
  \multicolumn{1}{c|}{80.60} &
  0.005549 &
  \multicolumn{1}{c|}{77.94} &
  0.4785 &
  \multicolumn{1}{c}{85.31}\\
  \multicolumn{1}{c|}{DCNv1 \cite{dcn}} &
  0.4376 &
  \multicolumn{1}{c|}{81.44} &
  0.3156 &
  \multicolumn{1}{c|}{90.38} &
  0.1501 &
  \multicolumn{1}{c|}{80.59} &
  0.005541 &
  \multicolumn{1}{c|}{78.13} &
  \underline{0.4766} &
  \multicolumn{1}{c}{85.31}\\
  \multicolumn{1}{c|}{xDeepFM \cite{xdeepfm}} &
  0.4376 &
  \multicolumn{1}{c|}{81.43} &
  0.3054 &
  \multicolumn{1}{c|}{90.47} &
  0.1501 &
  \multicolumn{1}{c|}{80.62} &
  0.005534 &
  \multicolumn{1}{c|}{78.25} &
  0.4772 &
  \multicolumn{1}{c}{85.35} \\
  \multicolumn{1}{c|}{AutoInt* \cite{autoint}} &
  0.4390 &
  \multicolumn{1}{c|}{81.32} &
  0.3112 &
  \multicolumn{1}{c|}{90.45} &
  0.1502 &
  \multicolumn{1}{c|}{80.57} &
  0.005544 &
  \multicolumn{1}{c|}{78.16} &
  0.4773 &
  \multicolumn{1}{c}{85.34} \\
  \multicolumn{1}{c|}{AFN* \cite{AFN}} &
  0.4384 &
  \multicolumn{1}{c|}{81.38} &
  0.3048 &
  \multicolumn{1}{c|}{90.53} &
   0.1499 &
  \multicolumn{1}{c|}{80.70} &
  0.005539 &
  \multicolumn{1}{c|}{78.17} &
  0.4842 &
  \multicolumn{1}{c}{84.89}\\
  \multicolumn{1}{c|}{DCNv2 \cite{dcnv2}} &
  0.4376 &
  \multicolumn{1}{c|}{81.45} &
  0.3098 &
  \multicolumn{1}{c|}{90.56} &
  0.1502 &
  \multicolumn{1}{c|}{80.59} &
  0.005539 &
  \multicolumn{1}{c|}{78.26} &
  0.4787 &
  \multicolumn{1}{c}{85.31}\\
  \multicolumn{1}{c|}{EDCN \cite{EDCN}} &
  0.4386 &
  \multicolumn{1}{c|}{81.36} &
  0.3073 &
  \multicolumn{1}{c|}{90.48} & 
  0.1501 &
  \multicolumn{1}{c|}{80.62} &
  0.005573 &
  \multicolumn{1}{c|}{77.93} & 
  0.4952 &
  \multicolumn{1}{c}{85.27} \\
  \multicolumn{1}{c|}{MaskNet \cite{masknet}} &
  0.4387 &
  \multicolumn{1}{c|}{81.34} &
  0.3080 &
  \multicolumn{1}{c|}{90.34} &
  0.1498 &
  \multicolumn{1}{c|}{80.79} &
  0.005556 &
  \multicolumn{1}{c|}{77.85} &
  0.5003 &
  \multicolumn{1}{c}{84.79}\\
  \multicolumn{1}{c|}{CL4CTR \cite{CL4CTR}} &
  0.4383 &
  \multicolumn{1}{c|}{81.35} &
  0.3074 &
  \multicolumn{1}{c|}{90.33} &
  0.1502 &
  \multicolumn{1}{c|}{80.56} &
  0.005543 &
  \multicolumn{1}{c|}{78.06} &
  0.4972  &
  \multicolumn{1}{c}{83.78}\\
  \multicolumn{1}{c|}{EulerNet \cite{EulerNet}} &
  0.4379 &
  \multicolumn{1}{c|}{81.47} &
  0.3050 &
  \multicolumn{1}{c|}{90.44} &
  0.1498 &
  \multicolumn{1}{c|}{80.78} &
  0.005540 &
  \multicolumn{1}{c|}{78.30} &
  0.4922 &
  \multicolumn{1}{c}{84.27}\\ 
  \multicolumn{1}{c|}{FinalMLP \cite{finalmlp}} &
  0.4373 &
  \multicolumn{1}{c|}{81.45} &
  0.3058 &
  \multicolumn{1}{c|}{90.52} &
  0.1497 &
  \multicolumn{1}{c|}{80.78} &
  0.005556 &
  \multicolumn{1}{c|}{78.02} &
  0.4822 &
  \multicolumn{1}{c}{85.10}\\ 
  \multicolumn{1}{c|}{FINAL(2B) \cite{FINAL}} &
  0.4371 &
  \multicolumn{1}{c|}{81.49} &
  0.3035 &
  \multicolumn{1}{c|}{90.53} &
  0.1498 &
  \multicolumn{1}{c|}{80.74} &
  0.005540 &
  \multicolumn{1}{c|}{78.13} &
  0.4800 &
  \multicolumn{1}{c}{85.14} \\
  \multicolumn{1}{c|}{RFM \cite{RFM}} &
  0.4374 &
  \multicolumn{1}{c|}{81.47} &
  0.3048 &
  \multicolumn{1}{c|}{90.51} &
  0.1506 &
  \multicolumn{1}{c|}{80.73} &
  0.005540 &
  \multicolumn{1}{c|}{78.25} &
  0.4853 &
  \multicolumn{1}{c}{84.70}\\ 
  \multicolumn{1}{c|}{ECN \cite{FCN}} &
  \underline{0.4364} &
  \multicolumn{1}{c|}{\underline{81.55}} &
  \underline{0.3013} &
  \multicolumn{1}{c|}{\underline{90.59}} &
  \underline{0.1496} &
  \multicolumn{1}{c|}{\underline{80.90}} &
  \underline{0.005534} &
  \multicolumn{1}{c|}{\underline{78.43}} &
  0.4778 &
  \multicolumn{1}{c}{\underline{85.40}} \\ \hline
  \multicolumn{1}{c|}{MLP \cite{DNN}} &
  0.4380 &
  \multicolumn{1}{c|}{81.40} & 
  0.3100 &
  \multicolumn{1}{c|}{90.30} &
  0.1502 &
  \multicolumn{1}{c|}{80.52} &
   0.005545 &
  \multicolumn{1}{c|}{78.06} &
  0.4811 &
  \multicolumn{1}{c}{85.01} \\ 
  \multicolumn{1}{c|}{\textit{Abs.} $\Updownarrow$ \textit{Imp}}  &
  -0.001 &
  \multicolumn{1}{c|}{+0.10} & 
  -0.006 &
  \multicolumn{1}{c|}{+0.31} &
  -0.0004 &
  \multicolumn{1}{c|}{+0.27} &
  -0.000007 &
  \multicolumn{1}{c|}{+0.26} &
  +0.0011 &
  \multicolumn{1}{c}{+0.38} \\ 
  \multicolumn{1}{c|}{$\textbf{KDEF}_{\text{MLP}_{b}}$} &
  0.4370 &
  \multicolumn{1}{c|}{81.50} & 
  0.3040 &
  \multicolumn{1}{c|}{\textbf{90.61}} &
  0.1498 &
  \multicolumn{1}{c|}{80.79} &
   0.005538 &
  \multicolumn{1}{c|}{78.32} &
  0.4822 &
  \multicolumn{1}{c}{85.39} \\ \hline
  \multicolumn{1}{c|}{$\textbf{KDEF}_{\text{MLP}}$} &
  \textbf{0.4360$^\star$} &
  \multicolumn{1}{c|}{\textbf{81.60$^\star$}} &
  0.3036 &
  \multicolumn{1}{c|}{\textbf{90.62}} &
  \underline{0.1496} &
  \multicolumn{1}{c|}{80.85} &
  0.005536 &
  \multicolumn{1}{c|}{78.32} &
  \textbf{0.4764$^\star$} &
  \multicolumn{1}{c}{\textbf{85.66$^\star$}} \\
  \multicolumn{1}{c|}{$\textbf{KDEF}_{\text{Lastest}}$} &
  \textbf{0.4356$^\star$} &
  \multicolumn{1}{c|}{\textbf{81.63$^\star$}} &
   \textbf{0.3001$^\star$} &
  \multicolumn{1}{c|}{\textbf{90.75$^\star$}} &
   \textbf{0.1494$^\star$} &
  \multicolumn{1}{c|}{\textbf{80.99$^\star$}} &
  \textbf{0.005530$^\star$} &
  \multicolumn{1}{c|}{\textbf{78.48$^\star$}} &
  \textbf{0.4742$^\star$} &
  \multicolumn{1}{c}{\textbf{85.67$^\star$}} \\ \hline
  \Xhline{1px}
\end{tabular}}
\vspace{-1em}
\label{implicit}
\end{table*}

To validate the effectiveness of KDEF, we introduce both homogeneous ($\textbf{KDEF}_{\text{MLP}}$) and heterogeneous networks ($\textbf{KDEF}_{\text{Latest}}$) into KDEF as student models, and further compare the performance of these two ensemble methods with several SOTA. $\textbf{KDEF}_{\text{MLP}_{b}}$ denotes the best student model performance in $\textbf{KDEF}_{\text{MLP}}$. \textit{Abs.Imp} represents the absolute performance improvement. The experimental results are shown in Table \ref{baselines}, where bold numbers indicate performance surpassing the baseline, and underlined values represent the best baseline performance. We can draw the following conclusions:
\begin{itemize}[leftmargin=*]
\item CTR researchers generally consider MLP to struggle with learning multiplicative feature interactions, leading to inherent performance limitations \cite{neuralvsmf, FINAL, CETN}. However, $\textbf{KDEF}_{\text{MLP}}$ surpasses all baselines on the Criteo and KKBox, and achieves performance comparable to SOTA on other datasets. Therefore, these results demonstrate the effectiveness of KDEF.
\item $\textbf{KDEF}_{\text{MLP}_{b}}$ shows AUC performance improvements over individually trained MLP on all five datasets, with each surpassing the standard threshold of \textit{0.1\%}. The performance gain is particularly significant on the KKBox dataset, where AUC improves by 0.38\%. This demonstrates that KDEF not only enhances the performance of the ensemble model but also improves the performance of individual sub-networks, ensuring model flexibility. In practical applications, when higher model performance is required, the ensemble model can be selected; conversely, when lower inference latency is needed, the best sub-network can be used. Notably, we find that knowledge transfer methods sometimes negatively impact Logloss optimization, which will be further explained in Section \ref{baseline_KT}.
\item $\textbf{KDEF}_{\text{Latest}}$ achieves SOTA performance across all datasets, consistently surpassing the selected sub-networks (i.e., PNN, FINAL(2B), ECN), further demonstrating the effectiveness of KDEF. Notably, KDEF performs well on the widely-used, highly sparse ($\textgreater$ 99.99\% \cite{autoint}) large-scale dataset, Criteo, achieving the best performance. Moreover, compared to the ensemble of multiple MLPs in $\textbf{KDEF}_{\text{MLP}}$, $\textbf{KDEF}_{\text{Latest}}$ achieves SOTA performance using only three heterogeneous CTR models, indicating that the ensemble of heterogeneous networks outperforms that of homogeneous networks. We believe that further integrating more heterogeneous CTR models within the KDEF could lead to even better performance, but this is beyond the scope of this paper, so we include it in future work.
\end{itemize}

\begin{table}[t]
\Huge
\renewcommand\arraystretch{1.2}
\centering
\caption{\textbf{Performance comparison of different knowledge transfer methods.}} 
\resizebox{\linewidth}{!}{
\begin{tabular}{c|cc|cc|cc}
\Xhline{2px}
\multirow{2}{*}{\textbf{Method}}  & \multicolumn{2}{c|}{\textbf{Criteo}} & \multicolumn{2}{c|}{\textbf{ML-1M}} & \multicolumn{2}{c}{\textbf{KKBox}}\\ \cline{2-7} 
                       & $ \text{Logloss}\downarrow$       & $ \text{AUC(\%)}\uparrow$        & $ \text{Logloss}\downarrow$       & $ \text{AUC(\%)}\uparrow$   &
                       $ \text{Logloss}\downarrow$       & $ \text{AUC(\%)}\uparrow$ \\ \hline
 \multicolumn{1}{c|}{Vanilla}  & 0.4380 & 81.40 & 0.3100 & 90.30  & \underline{0.4811} & 85.01  \\ \hline
 \multicolumn{1}{c|}{Online KD \cite{KD}}  & 0.4379 & 81.41 & \underline{0.3083} & \underline{90.40}  & 0.4848 & 85.16  \\
 \multicolumn{1}{c|}{KDCL \cite{KDCL}}  & \underline{0.4375} & \underline{81.46} & 0.3134 & 90.35  & 0.4857 & \underline{85.30} \\
  \multicolumn{1}{c|}{ECKD \cite{KD4CTR}}  & 0.4381 & 81.41 & 0.3112 & 90.35 & 0.4859 & 85.17 \\
  \multicolumn{1}{c|}{DAGFM \cite{DAGFM}}  & 0.4380 & 81.42 & 0.3096 & 90.37 & 0.4862 & 85.11 \\ \hline
  \multicolumn{1}{c|}{$\textbf{KDEF}_{\text{MLP}}$}  & \textbf{0.4360} & \textbf{81.60} & \textbf{0.3036} & \textbf{90.62}  & \textbf{0.4764} & \textbf{85.66} \\
  \Xhline{2px}
\end{tabular}}
\label{KT}
\end{table}

\subsubsection{\textbf{Performance Comparison with Other Knowledge Transfer Methods (RQ3).}}
\label{baseline_KT}
To verify the superiority of KDEF over other knowledge transfer methods, we compare it with four methods based on knowledge transfer. To ensure a fair comparison, we adopt MLP as the backbone and keep all hyperparameters fixed, except for the loss balance parameters. Table \ref{KT} shows that KDEF outperforms all baseline methods, while KDCL and Online KD take second place in AUC. Meanwhile, we observe that while these knowledge transfer methods can improve AUC, they may sometimes lead to negative gains in Logloss optimization on the ML-1M and KKBox datasets. This could be due to the inherent bias in the soft labels generated by the knowledge transfer method, as compared to the true labels. Moreover, these knowledge transfer methods that introduce additional loss functions often come with high hyperparameter tuning costs, which is one of the reasons for their suboptimal performance. In contrast, KDEF is both effective and hyperparameter-free.

\begin{table}[t]
\Huge
\renewcommand\arraystretch{1.4}
\centering
\caption{\textbf{Ablation study of KDEF.}} 
\resizebox{\linewidth}{!}{
\begin{tabular}{c|cc|cc|cc|cc}
\Xhline{1px}
\multirow{2}{*}{\textbf{Model}}  & \multicolumn{2}{c|}{\textbf{Criteo}} & \multicolumn{2}{c|}{\textbf{ML-1M}} & \multicolumn{2}{c|}{\textbf{KKBox}} & \multicolumn{2}{c}{\textbf{iPinYou}}\\ \cline{2-9} 
                       & $ \text{Logloss}\downarrow$       & $ \text{AUC(\%)}\uparrow$        & $ \text{Logloss}\downarrow$       & $ \text{AUC(\%)}\uparrow$   &
                       $ \text{Logloss}\downarrow$       & $ \text{AUC(\%)}\uparrow$ &
                       $ \text{Logloss}\downarrow$       & $ \text{AUC(\%)}\uparrow$\\ \hline
\multicolumn{1}{c|}{PNN \cite{pnn1}}  & 0.4378 & 81.42 & 0.3070 & 90.42  & 0.4793 & 85.15 & 0.005544 & 78.13\\
\multicolumn{1}{c|}{$\textbf{KDEF}_{\text{PNN}}$}  &\textbf{0.4365} & \textbf{81.54} & \textbf{0.3039} & \textbf{90.67}  & \textbf{0.4769} & \textbf{85.54} & \textbf{0.005549} & \textbf{78.19} \\ \hline
\multicolumn{1}{c|}{FINAL (1B) \cite{FINAL}}  & 0.4377 & 81.44 & 0.3053 & 90.41 & 0.4830 & 85.13 & 0.005541 & 78.10 \\
\multicolumn{1}{c|}{$\textbf{KDEF}_{\text{FINAL(1B)}}$}  & \textbf{0.4364} & \textbf{81.55} & \textbf{0.3041} & \textbf{90.64}  & \textbf{0.4776} & \textbf{85.55} & \textbf{0.005542} & \textbf{78.38} \\ \hline
\multicolumn{1}{c|}{ECN \cite{FCN}}  & 0.4364 & 81.55 & 0.3013 & 90.59  & 0.4778 & 85.40 & 0.005534 & 78.43\\ 
\multicolumn{1}{c|}{$\textbf{KDEF}_{\text{ECN}}$}  & \textbf{0.4362} & \textbf{81.58} & \textbf{0.2989} & \textbf{90.65}  & \textbf{0.4757} & \textbf{85.51} & \textbf{0.005541} & \textbf{78.36} \\ \hline
only KD         & 0.4369   & 81.53  & 0.3055 & 90.61 & {0.4822} & {85.58}  & 0.005538 & 78.28  \\ 
only DML        & 0.4369  & 81.51 & 0.3059  & 90.50 & 0.4968 & 85.02 & 0.005544 & 78.22  \\
w/o EM        & {0.4360} & {81.60} & {0.3012} & {90.65} & 0.4763 & 85.48 & 0.005537 & 78.32 \\ 
w/o all        & 0.4383 & 81.38 & 0.3075 & 90.51 & 0.4842 & 85.33 & 0.005534 & 78.20 \\ \hline 
\multicolumn{1}{c|}{$\textbf{KDEF}_{\text{Lastest}}$}  & \textbf{0.4356} & \textbf{81.63} & \textbf{0.3001} & \textbf{90.75}  & \textbf{0.4742} & \textbf{85.67} & \textbf{0.005530} & \textbf{78.48} \\ 
 \Xhline{1px}
\end{tabular}}
\label{ablation}
\end{table}

\subsection{In-Depth Study of KDEF}
\subsubsection{\textbf{Ablation Study (RQ4).}}
To investigate the impact of each component of KDEF on its performance, we conduct experiments on several variants of KDEF:
\begin{itemize}[leftmargin=*]
\item \textbf{only KD}: $\textbf{KDEF}_{\text{Latest}}$ using only knowledge distillation.
\item \textbf{only DML}: $\textbf{KDEF}_{\text{Latest}}$ using only deep mutual learning.
\item \textbf{w/o EM}: $\textbf{KDEF}_{\text{Latest}}$ without the examination mechanism. Following prior works \cite{FINAL, FCN} that introduce auxiliary loss functions, we set $\lambda = \mu = 1$.
\item \textbf{w/o all}: The ensemble is performed using only the mean operation.
\end{itemize}
From Table \ref{ablation}, we observe that all variants of KDEF have some degree of performance degradation, demonstrating the necessity of each component. Notably, using \textbf{only DML} for sub-network ensemble can sometimes decrease performance. For instance, on the KKBox dataset, the \textbf{only DML} variant performs worse than the \textbf{w/o all} variant, a result similar to that reported in \cite{DML4CTR}. The \textbf{w/o EM} variant exhibits minor performance degradation on the Criteo dataset but suffers greater losses on ML-1M, KKBox, and iPinYou. Combined with the results in Table \ref{baselines}, we observe that this is because the performance gap among PNN, FINAL, and ECN is relatively small on Criteo, but significantly larger on the other datasets. This suggests that selective learning between teacher-to-student and peer-to-peer modes becomes less critical when the performance disparity among student networks is small. In such cases, applying a unified loss balancing weight is sufficient, which explains why \textbf{w/o EM} does not work well on Criteo. Moreover, we observe that not only does $\textbf{KDEF}_{\text{Latest}}$ achieve better performance, but its corresponding sub-networks, $\textbf{KDEF}_{\text{PNN}}$, $\textbf{KDEF}_{\text{FINAL(1B)}}$, and $\textbf{KDEF}_{\text{ECN}}$, also achieve noticeable performance improvements. This demonstrates that KDEF can enhance both the ensemble performance and the performance of individual sub-networks. Meanwhile, by comparing PNN, FINAL, ECN, and the w/o all variant, we find that simply applying average ensemble to multiple sub-networks often leads to performance degradation, which is consistent with our findings in Section \ref{section 3}.

\subsubsection{\textbf{Impact of Different Loss (RQ4)}}
\begin{figure}[t]
    \centering
    \begin{minipage}[t]{0.33\linewidth}
        \centering
        \includegraphics[width=\linewidth]{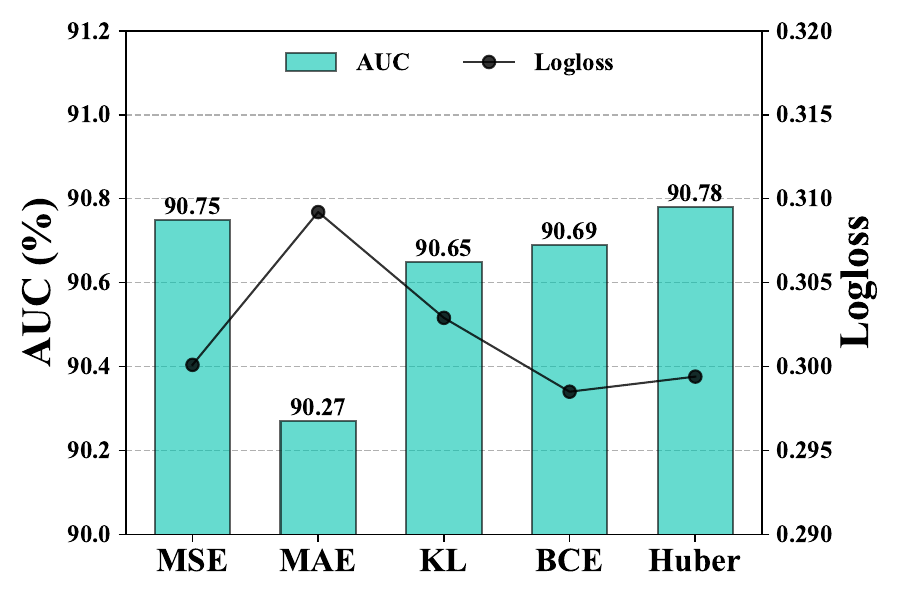}
        \subcaption{\footnotesize ML-1M}
    \end{minipage}%
    \hspace{-0.2em} 
    \begin{minipage}[t]{0.33\linewidth}
        \centering
        \includegraphics[width=\linewidth]{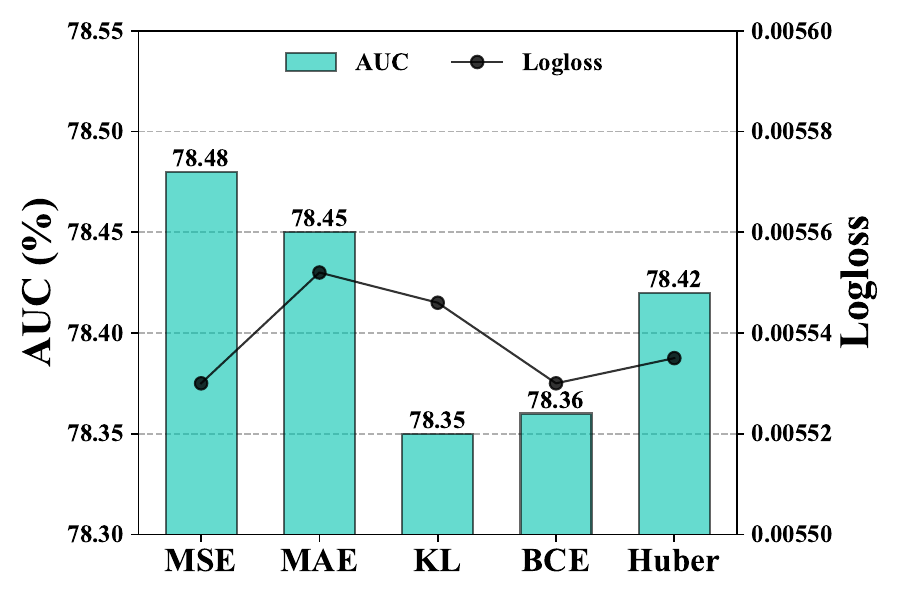}
        \subcaption{\footnotesize iPinYou}
    \end{minipage}%
    \hspace{-0.2em} 
    \begin{minipage}[t]{0.33\linewidth}
        \centering
        \includegraphics[width=\linewidth]{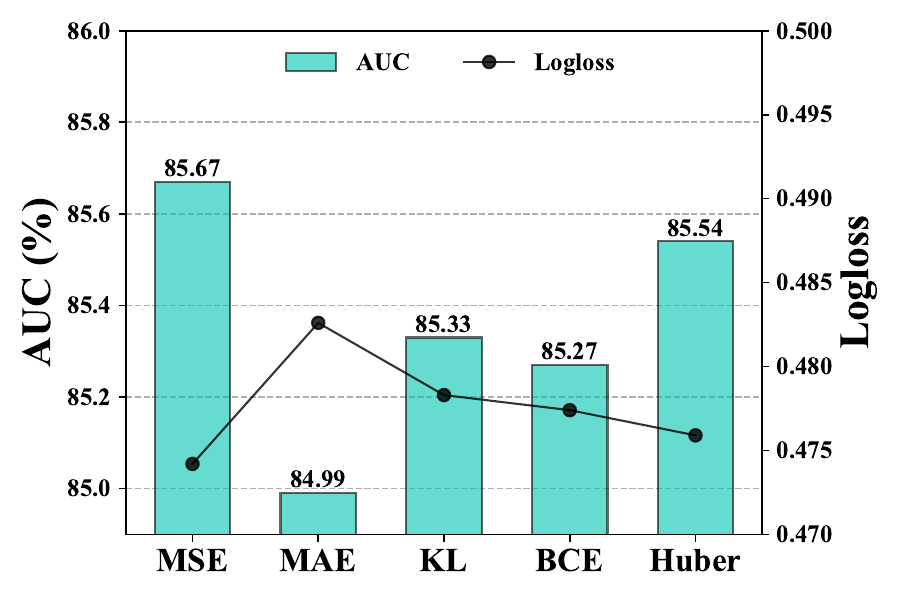}
        \subcaption{\footnotesize KKBox}
    \end{minipage}
    \captionsetup{justification=raggedright}
    \caption{Performance of different Loss in KDEF$_{\text{Lastest}}$.}
    \label{Loss}
\end{figure}

To investigate the impact of different consistency losses on KDEF, we replace the MSE loss with various alternatives. The experimental results, as shown in Fig. \ref{Loss}, indicate that MSE Loss consistently demonstrates superior performance across all three datasets, with only a slight disadvantage compared to Huber Loss on the ML-1M dataset. Moreover, we find that the KL divergence loss, which is widely used in the KD, performs relatively poorly. This may be because KL divergence is more sensitive to the extreme values and sparsity of probability distributions, which can adversely affect model optimization.

\begin{table}[t]
\Huge
\renewcommand\arraystretch{1.2}
\centering
\caption{\textbf{Compatibility study of KDEF.}} 
\resizebox{\linewidth}{!}{
\begin{tabular}{c|cc|cc|cc}
\Xhline{2px}
\multirow{2}{*}{\textbf{Model}}  & \multicolumn{2}{c|}{\textbf{Criteo}} & \multicolumn{2}{c|}{\textbf{ML-1M}} & \multicolumn{2}{c}{\textbf{KKBox}}\\ \cline{2-7} 
                       & $ \text{Logloss}\downarrow$       & $ \text{AUC(\%)}\uparrow$        & $ \text{Logloss}\downarrow$       & $ \text{AUC(\%)}\uparrow$   &
                       $ \text{Logloss}\downarrow$       & $ \text{AUC(\%)}\uparrow$ \\ \hline
   \multicolumn{1}{c|}{CrossNetv2 \cite{dcnv2}}  & 0.4392 & 81.27 & 0.3420 & 87.40  & 0.4912 & 84.42     \\
  \multicolumn{1}{c|}{$\textbf{KDEF}_{\text{CrossNetv2}}$}  & \textbf{0.4365} & \textbf{81.57} & \textbf{0.3021} & \textbf{90.45}  & \textbf{0.4865} & \textbf{84.69} \\ \hline
  \multicolumn{1}{c|}{DCNv2 \cite{dcnv2}}  & 0.4376 & 81.45 & 0.3098 & 90.56  & 0.4787 & 85.31     \\
  \multicolumn{1}{c|}{$\textbf{KDEF}_{\text{DCNv2}}$}  & \textbf{0.4366} & \textbf{81.56} & \textbf{0.3031} & \textbf{90.62}  & \textbf{0.4780} & \textbf{85.56} \\ \hline
  \multicolumn{1}{c|}{CIN \cite{xdeepfm}}  & 0.4393 & 81.27 &  0.3432 & 87.32  & 0.4907 & 84.27     \\
  \multicolumn{1}{c|}{$\textbf{KDEF}_{\text{CIN}}$}  & \textbf{0.4385} & \textbf{81.36} & \textbf{0.3013} & \textbf{90.59}  & \textbf{0.4897} & \textbf{84.58} \\ \hline
    \multicolumn{1}{c|}{xDeepFM \cite{xdeepfm}}  & 0.4376 & 81.43 & 0.3054 & 90.47  & 0.4772 & 85.35     \\
  \multicolumn{1}{c|}{$\textbf{KDEF}_{\text{xDeepFM}}$}  & \textbf{0.4369} & \textbf{81.51} & \textbf{0.3032} & \textbf{90.60}  & \textbf{0.4778} & \textbf{85.57} \\ \hline
  \multicolumn{1}{c|}{Wide \& Deep \cite{widedeep}}  & 0.4376 & 81.42 & 0.3056 & 90.45 & 0.4852 & 85.04     \\
  \multicolumn{1}{c|}{$\textbf{KDEF}_{\text{Wide \& Deep}}$}  & \textbf{0.4368} & \textbf{81.52} & \textbf{0.3036} & \textbf{90.58}  & \textbf{0.4790} & \textbf{85.41} \\ \hline
    \multicolumn{1}{c|}{AFN* \cite{AFN}}  & 0.4384 & 81.38 & 0.3048 & 90.53  & 0.4842 & 84.89     \\ 
  \multicolumn{1}{c|}{$\textbf{KDEF}_{\text{AFN*}}$}  & \textbf{0.4372} & \textbf{81.49} & \textbf{0.3031} & \textbf{90.58}  & \textbf{0.4788} & \textbf{85.49} \\ \hline
  \multicolumn{1}{c|}{AutoInt* \cite{autoint}}  & 0.4390 & 81.32 & 0.3112 & 90.45  & 0.4773 & 85.34     \\
  \multicolumn{1}{c|}{$\textbf{KDEF}_{\text{AutoInt*}}$}  & \textbf{0.4368} & \textbf{81.52} & \textbf{0.3035} & \textbf{90.56}  & \textbf{0.4776} & \textbf{85.46} \\ \hline
  \Xhline{2px}
\end{tabular}}
\label{Compatibility}
\end{table}

\subsubsection{\textbf{Compatibility  Study (RQ5)}}
We propose KDEF as a model-agnostic and hyperparameter-free training framework. To verify whether KDEF can replace traditional CTR training paradigms and generalize to more models, we further evaluate the compatibility and flexibility of the KDEF ensemble training method. We use the ensemble setup of FINAL + ECN + X, where X is the model to be optimized. The experimental results are shown in Table \ref{Compatibility}. KDEF can directly use a single network or an assembled CTR model as its underlying sub-network. It is observed that under the influence of KDEF, CrossNetv2 without DNN integration achieves the best performance on the Criteo and KKBox datasets, particularly achieving a 0.3\% absolute gain on the Criteo dataset. This demonstrates that KDEF not only significantly improves the performance of ensemble models but also effectively enhances the performance of various sub-networks. When higher performance is required in production environments, the KDEF-enhanced ensemble model can be selected. Meanwhile, in scenarios with limited computational resources, the sub-network with the best performance can be directly used. This characteristic highlights the compatibility and flexibility of KDEF.

\begin{table}[t]
\Huge
\renewcommand\arraystretch{1.3}
\centering
\caption{\textbf{Efficiency Study of KDEF.}} 
\resizebox{\linewidth}{!}{
\begin{tabular}{c|c|c|ccc|ccc}
\Xhline{2px}
\multirow{2}{*}{Model} & \multirow{2}{*}{\#Params} & \multirow{2}{*}{\#Time×Epochs} & \multicolumn{3}{c|}{Ensemble} & \multicolumn{3}{c}{Best sub-network} \\ \cline{4-9} 
          &        &           & Logloss & AUC   & \#Latency & Logloss & AUC   & \#Latency \\ \hline
$\textbf{KDEF}_{\text{MLP x2}}$         & 15.71M & 4min×28   & 0.4368  & 81.52 & 0.21ms    & 0.4374  & 81.47 & 0.14ms    \\
$\textbf{KDEF}_{\text{MLP x3}}$         & 16.29M & 4.5min×19 & 0.4366  & 81.53 & 0.26ms    & 0.4373  & 81.48 & 0.15ms         \\
$\textbf{KDEF}_{\text{MLP x4}}$         & 16.86M & 5min×21   & 0.4364  & 81.56 & 0.28ms    & 0.4371  & 81.49 & 0.14ms         \\
$\textbf{KDEF}_{\text{MLP x5}}$         & 17.42M & 5.5min×23 & 0.4362  & 81.58 & 0.33ms    & \underline{0.4370}  & \underline{81.50} & 0.17ms         \\
$\textbf{KDEF}_{\text{MLP x6}}$         & 18.01M & 6min×17   & 0.4364  & 81.57 & 0.36ms    & 0.4372  & 81.49 & 0.15ms         \\
$\textbf{KDEF}_{\text{MLP x10}}$        & 20.30M & 7min×28   & \underline{0.4360}  & \underline{81.60} & 0.55ms    & \underline{0.4370}  & \underline{81.50} & 0.16ms         \\ \hline
$\textbf{KDEF}_{\text{Lastest}}$   & 17.05M & 5min×20   & \textbf{0.4356}  & \textbf{81.63} & 1.57ms    & \textbf{0.4362}  & \textbf{81.58} & 0.39ms    \\ \hline
FINAL(2B) & 18.15M & 4min×15   & 0.4371  & 81.49 & 0.92ms    & -       & -     & -         \\ \Xhline{2px}
\end{tabular}}
\label{efficiency}
\end{table}

\subsubsection{\textbf{Efficiency and Flexibility Study (RQ5)}}
To evaluate the training and inference efficiency of KDEF, we conduct experiments on the Criteo dataset and compare it with the FINAL(2B) model deployed in industrial production \cite{FINAL}. As shown in Table \ref{efficiency}, the performance of both the ensemble model and the best student network of $\textbf{KDEF}_{\text{MLP}}$ improves progressively as the number of sub-networks increases, while the total number of parameters remains comparable to FINAL(2B). Due to the excellent parallelization efficiency of MLP, even with a tenfold increase in the number of sub-networks, the inference latency of the ensemble model remains lower than that of FINAL(2B). When the number of sub-networks is 5, the best student network outperforms FINAL(2B) while achieving lower inference latency. For $\textbf{KDEF}_{\text{Lastest}}$, although its ensemble model has higher inference latency, it achieves SOTA performance. Moreover, its best-performing sub-network sacrifices a small amount of performance but achieves nearly a fourfold improvement in inference speed. This further demonstrates the flexibility and efficiency of KDEF. In production environments, the ensemble model can be selected for higher performance, while the best student network can be used to reduce inference latency.

\subsection{Does KDEF Makes Sub-network More Similar?}
In the previous section, we have demonstrated the effectiveness of the KDEF. However, from a theoretical perspective, we observe that $\mathcal{L}_{KD}$ provides identical posteriors for the sub-networks, while $\mathcal{L}_{DML}$ provides similar posteriors. This raises several interesting questions: Are the representations learned by different sub-networks highly similar, especially when the sub-networks share a homogeneous architecture? Can KDEF mitigate the issue of information redundancy caused by overly similar representations?

\begin{figure}[t]
    \centering
    \begin{minipage}[t]{0.33\linewidth}
        \centering
        \includegraphics[width=\linewidth]{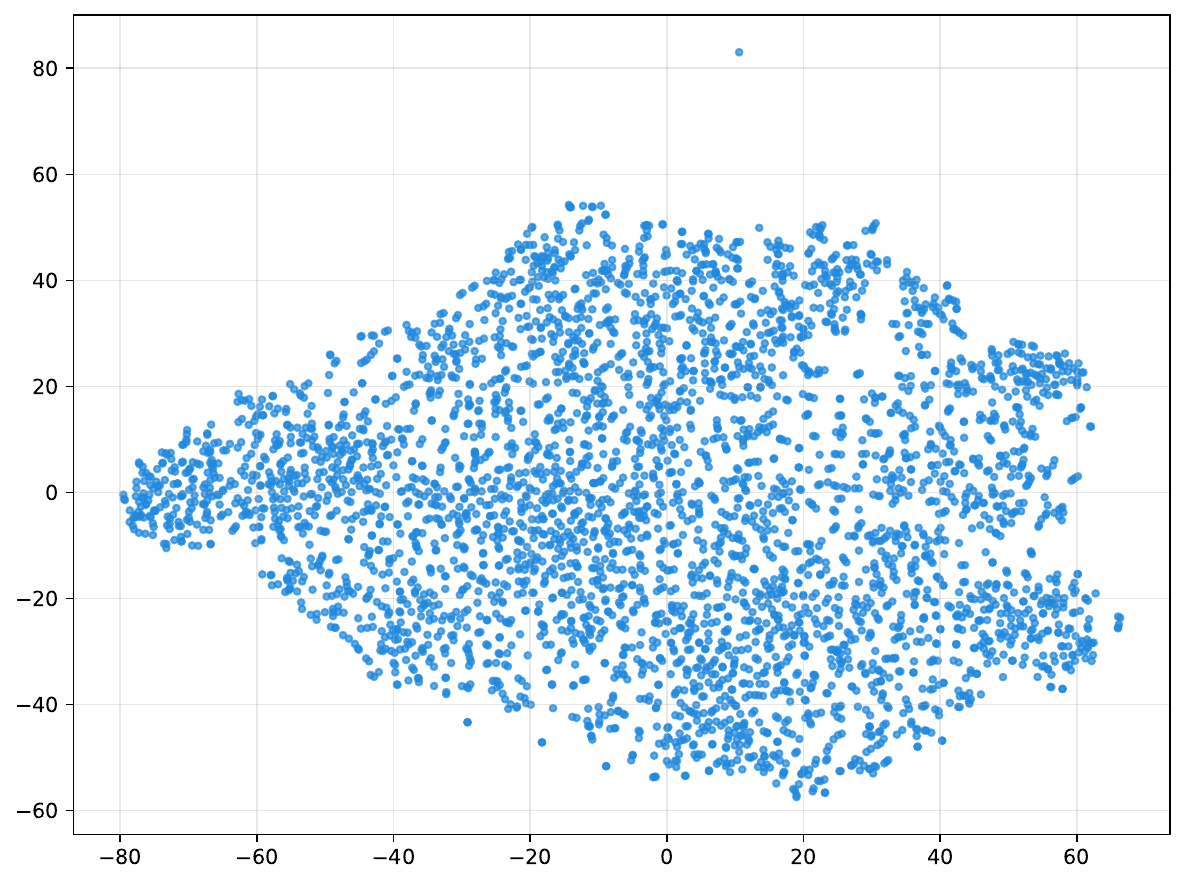}
        \subcaption{\footnotesize MLP$_1$}
    \end{minipage}
    \hspace{-0.5em} 
    \begin{minipage}[t]{0.33\linewidth}
        \centering
        \includegraphics[width=\linewidth]
        {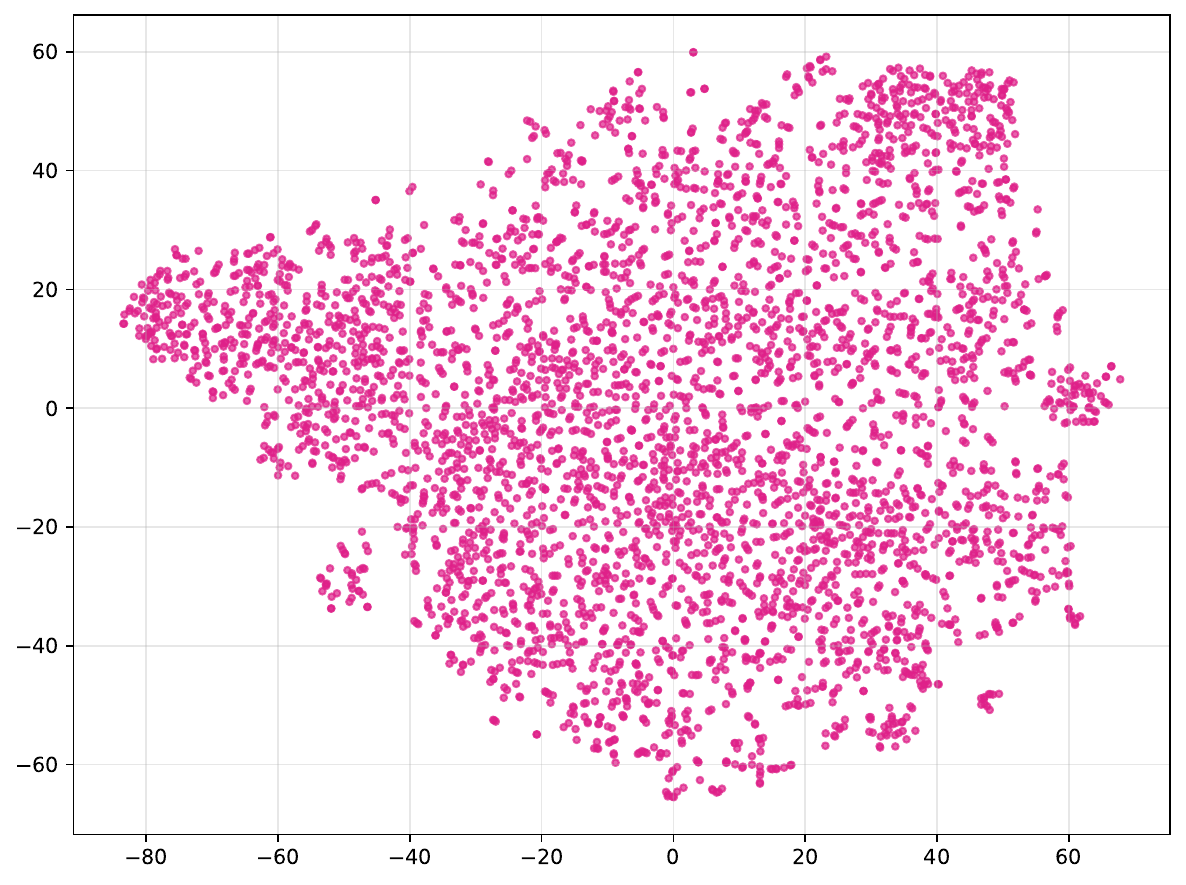}
        \subcaption{\footnotesize MLP$_2$}
    \end{minipage}
    \hspace{-0.5em} 
    \begin{minipage}[t]{0.33\linewidth}
        \centering
        \includegraphics[width=\linewidth]{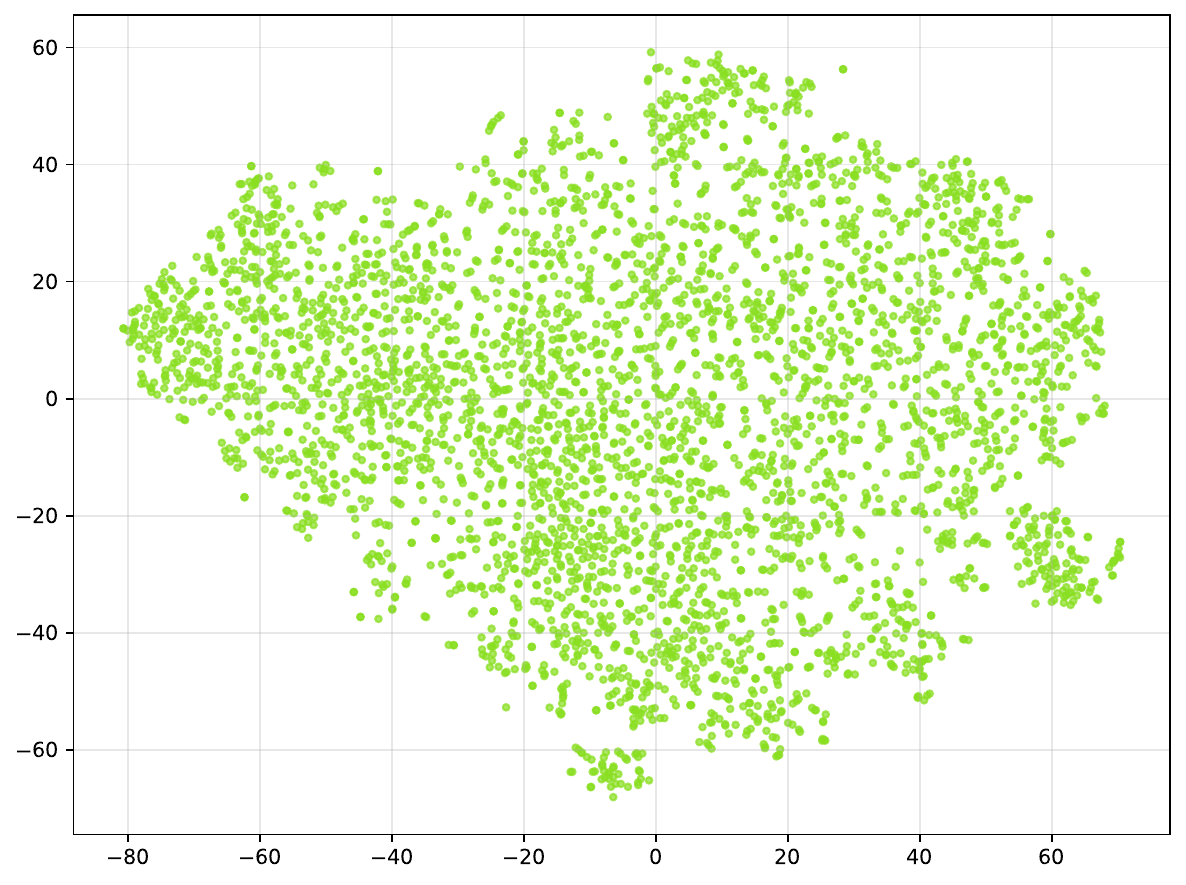}
        \subcaption{\footnotesize MLP$_3$}
    \end{minipage}%
    
    \begin{minipage}[t]{0.33\linewidth}
        \centering
        \includegraphics[width=\linewidth]{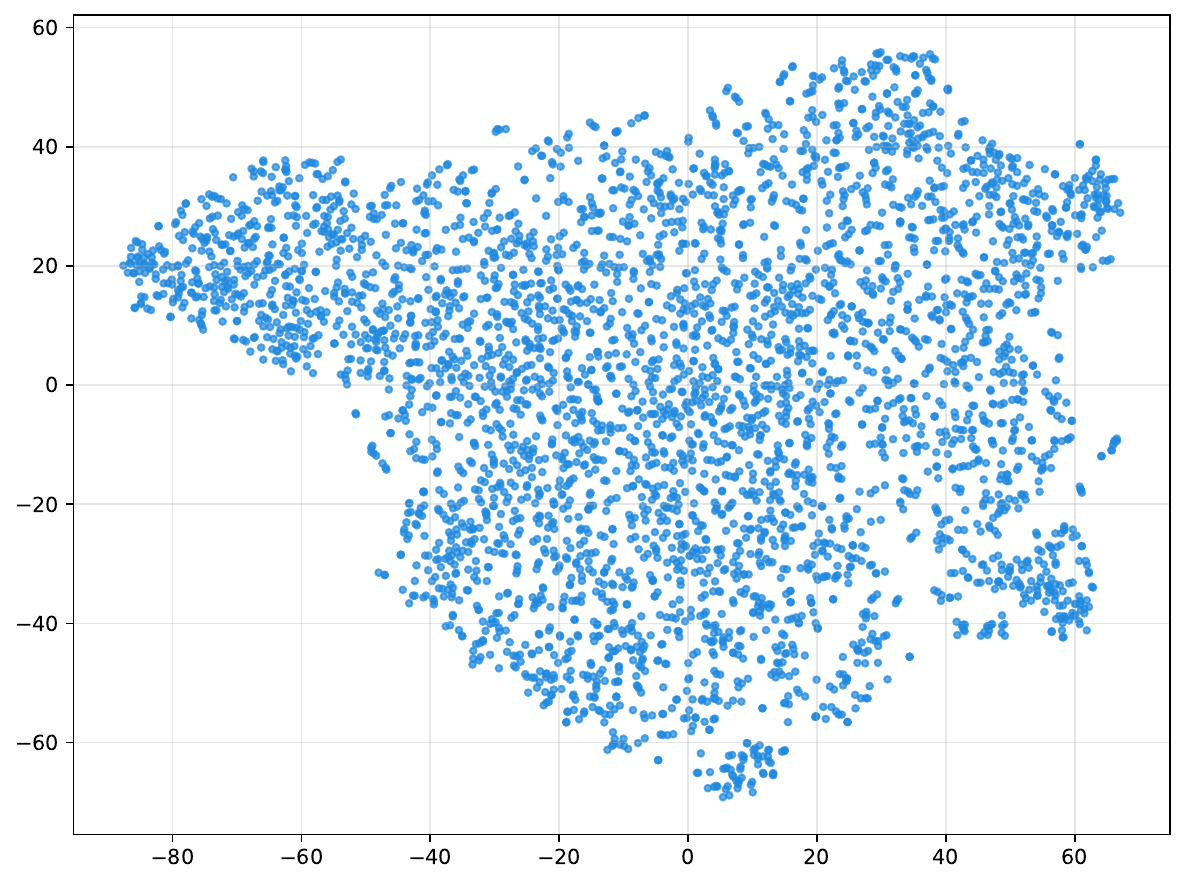}
        \subcaption{\footnotesize MLP$_1$ with KD}
    \end{minipage}
    \hspace{-0.5em} 
    \begin{minipage}[t]{0.33\linewidth}
        \centering
        \includegraphics[width=\linewidth]
        {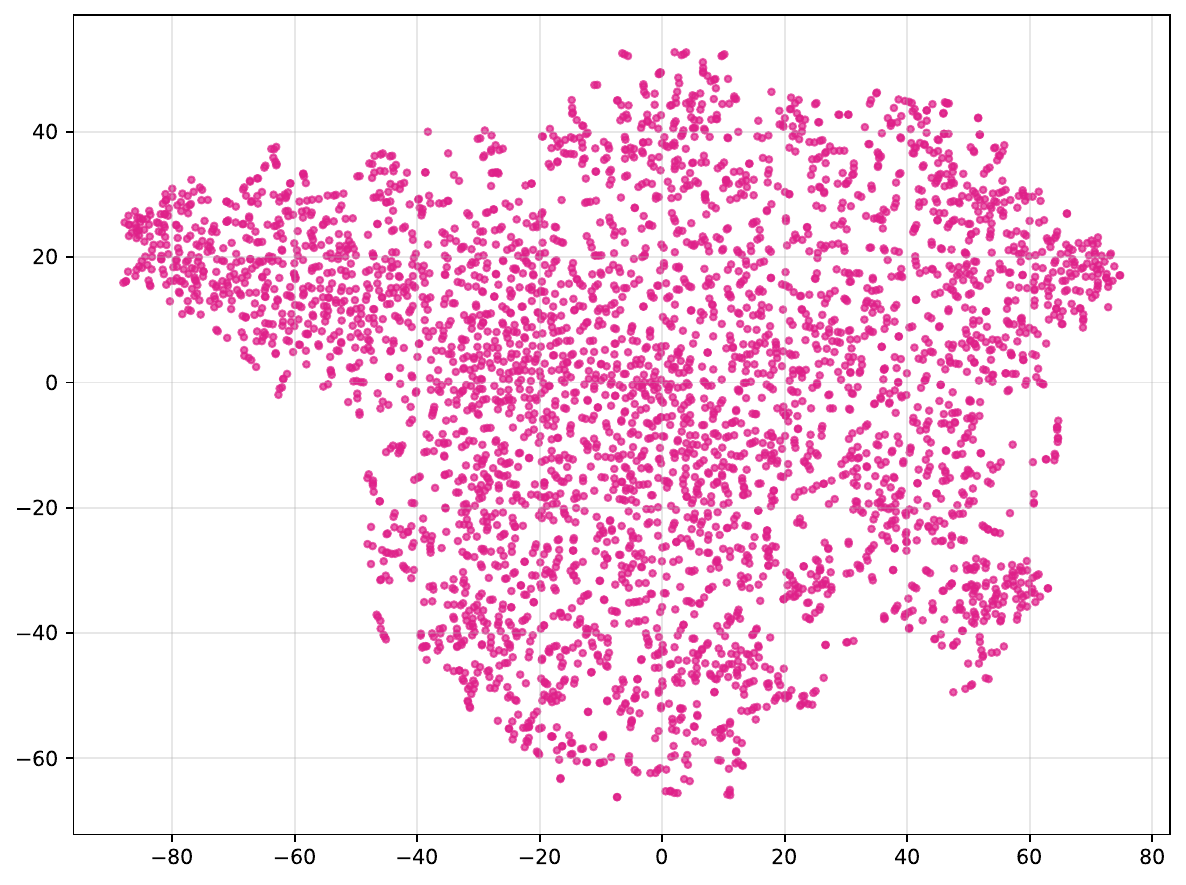}
        \subcaption{\footnotesize MLP$_2$ with KD}
    \end{minipage}
    \hspace{-0.5em} 
    \begin{minipage}[t]{0.33\linewidth}
        \centering
        \includegraphics[width=\linewidth]{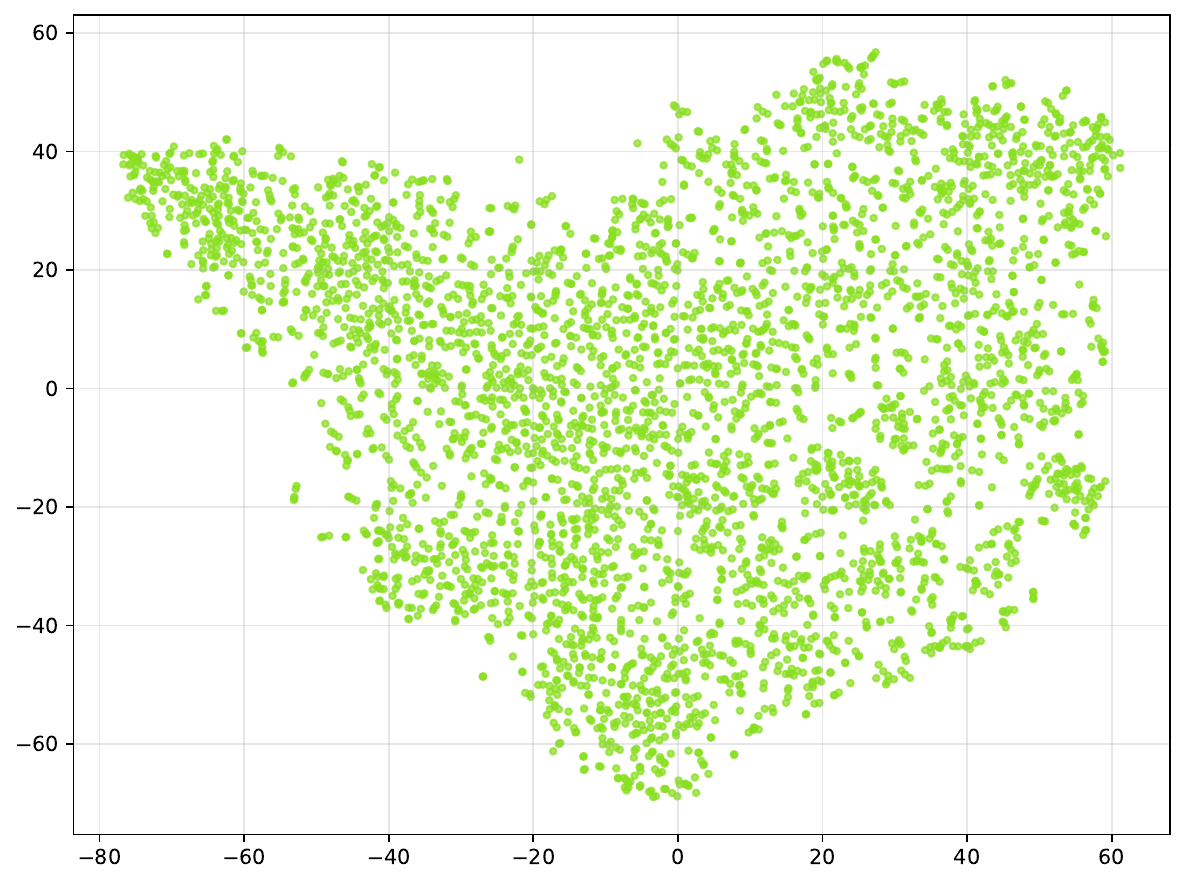}
        \subcaption{\footnotesize MLP$_3$ with KD}
    \end{minipage}%

    \begin{minipage}[t]{0.33\linewidth}
        \centering
        \includegraphics[width=\linewidth]{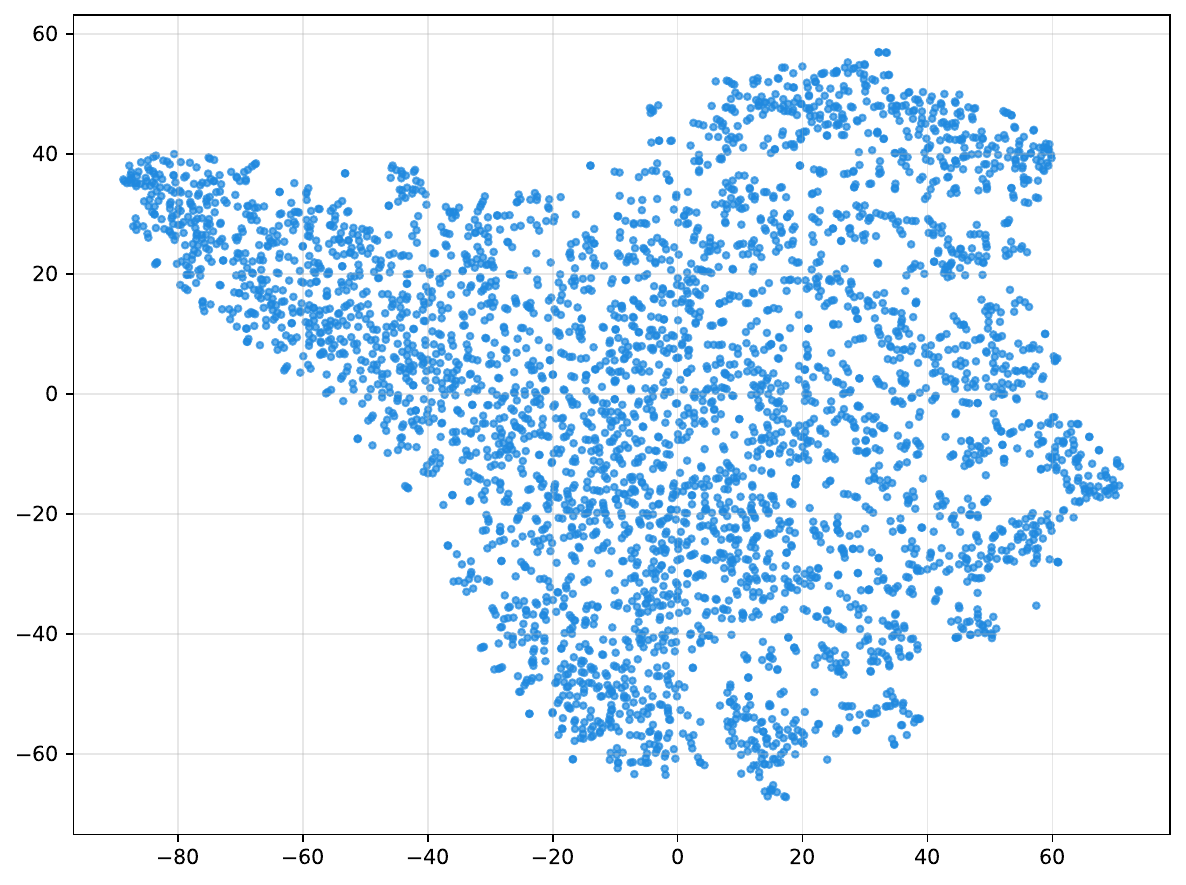}
        \subcaption{\footnotesize MLP$_1$ with DML}
    \end{minipage}
    \hspace{-0.5em} 
    \begin{minipage}[t]{0.33\linewidth}
        \centering
        \includegraphics[width=\linewidth]{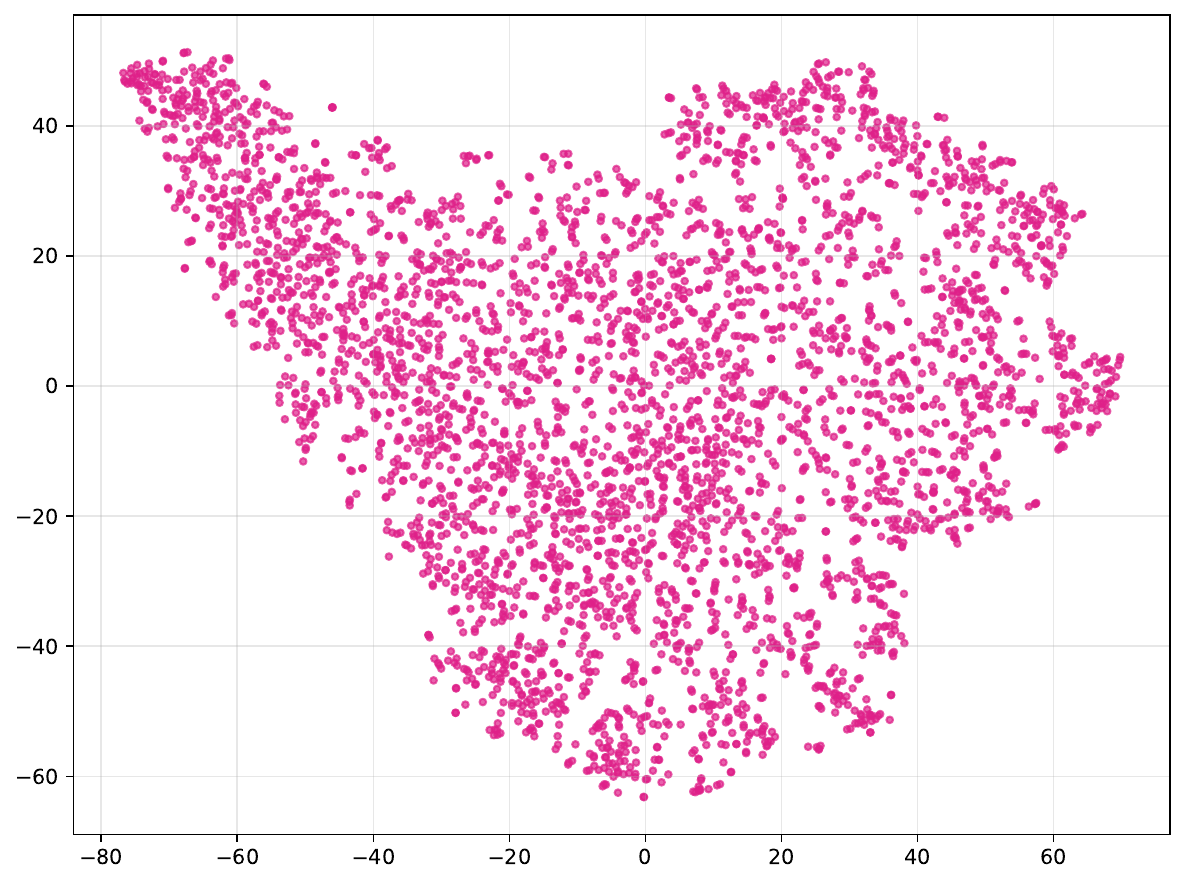}
        \subcaption{\footnotesize MLP$_2$ with DML}
    \end{minipage}
    \hspace{-0.5em} 
    \begin{minipage}[t]{0.33\linewidth}
        \centering
        \includegraphics[width=\linewidth]{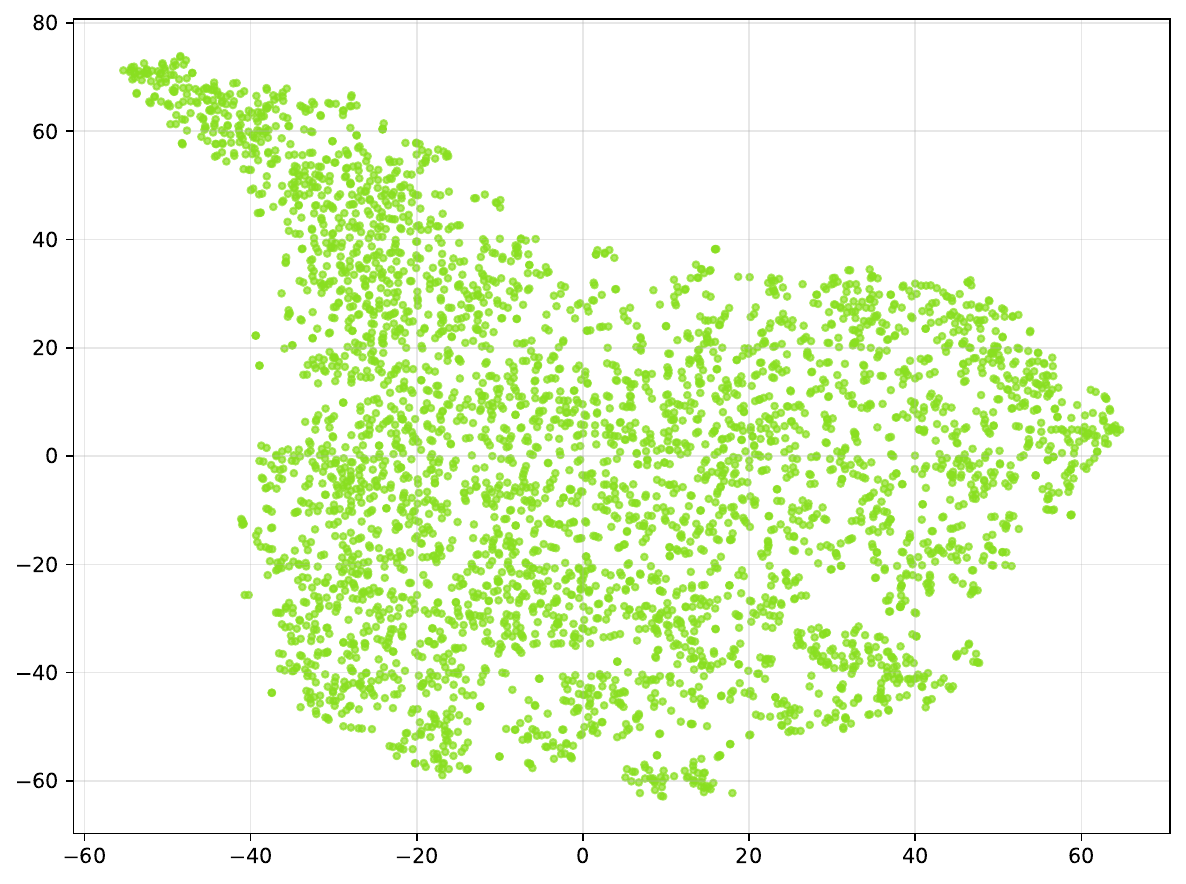}
        \subcaption{\footnotesize MLP$_3$ with DML}
    \end{minipage}%

    \begin{minipage}[t]{0.33\linewidth}
        \centering
        \includegraphics[width=\linewidth]{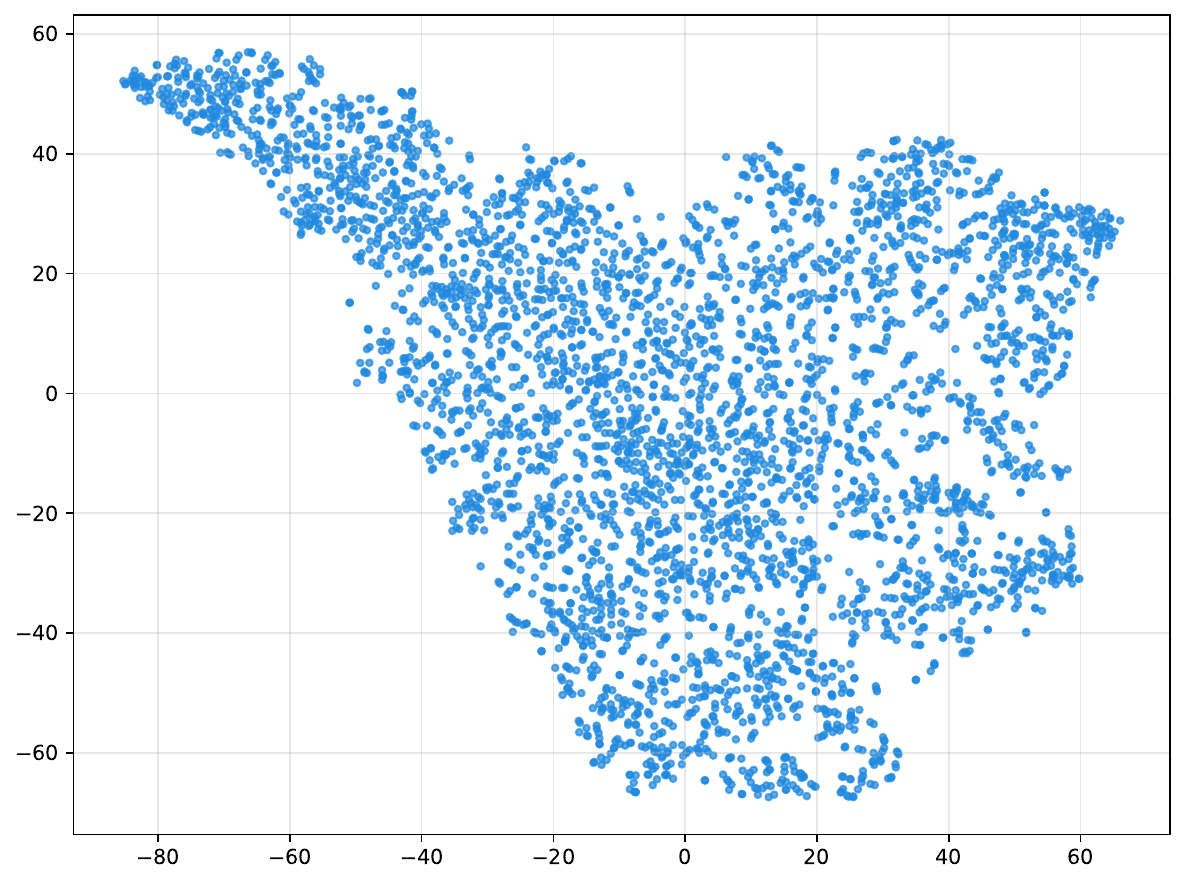}
        \subcaption{\footnotesize MLP$_1$ with KDEF}
    \end{minipage}
    \hspace{-0.5em} 
    \begin{minipage}[t]{0.33\linewidth}
        \centering
        \includegraphics[width=\linewidth]{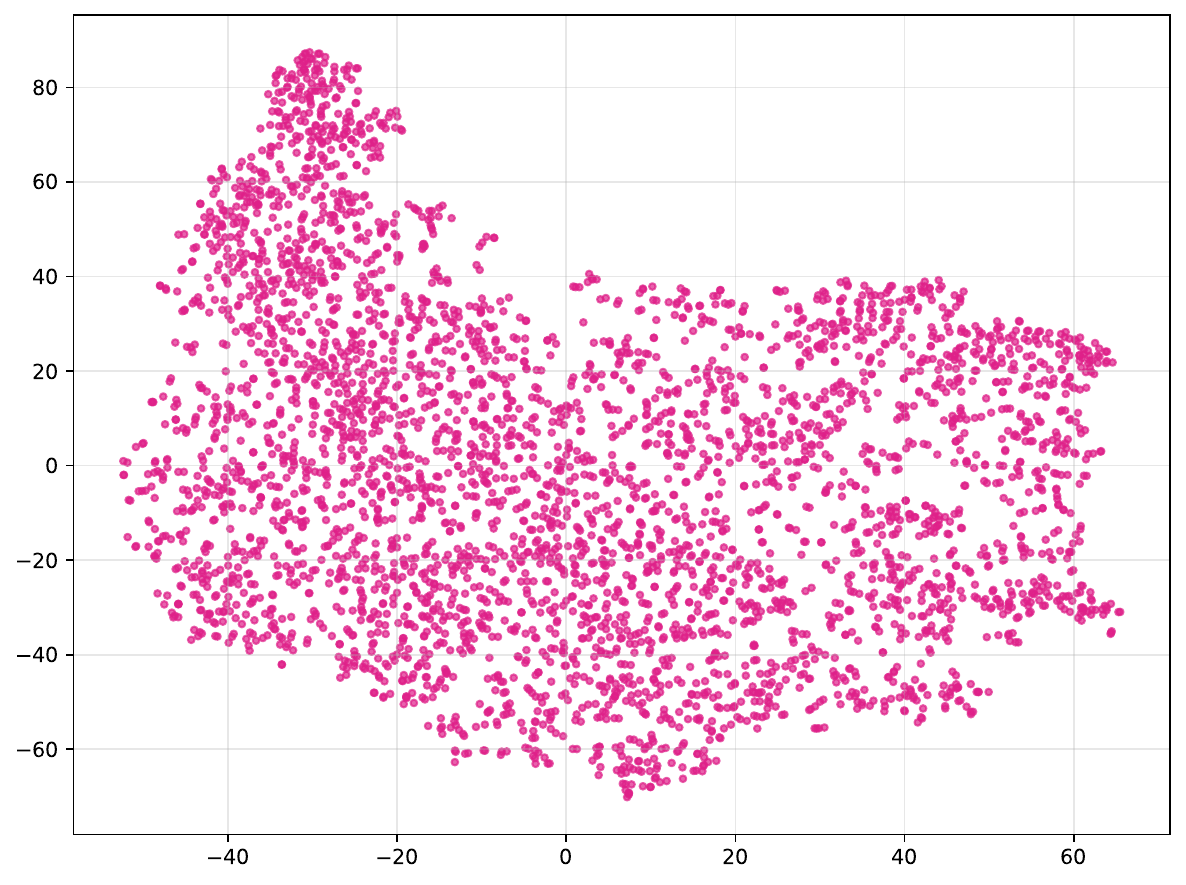}
        \subcaption{\footnotesize MLP$_2$ with KDEF}
    \end{minipage}
    \hspace{-0.5em} 
    \begin{minipage}[t]{0.33\linewidth}
        \centering
        \includegraphics[width=\linewidth]{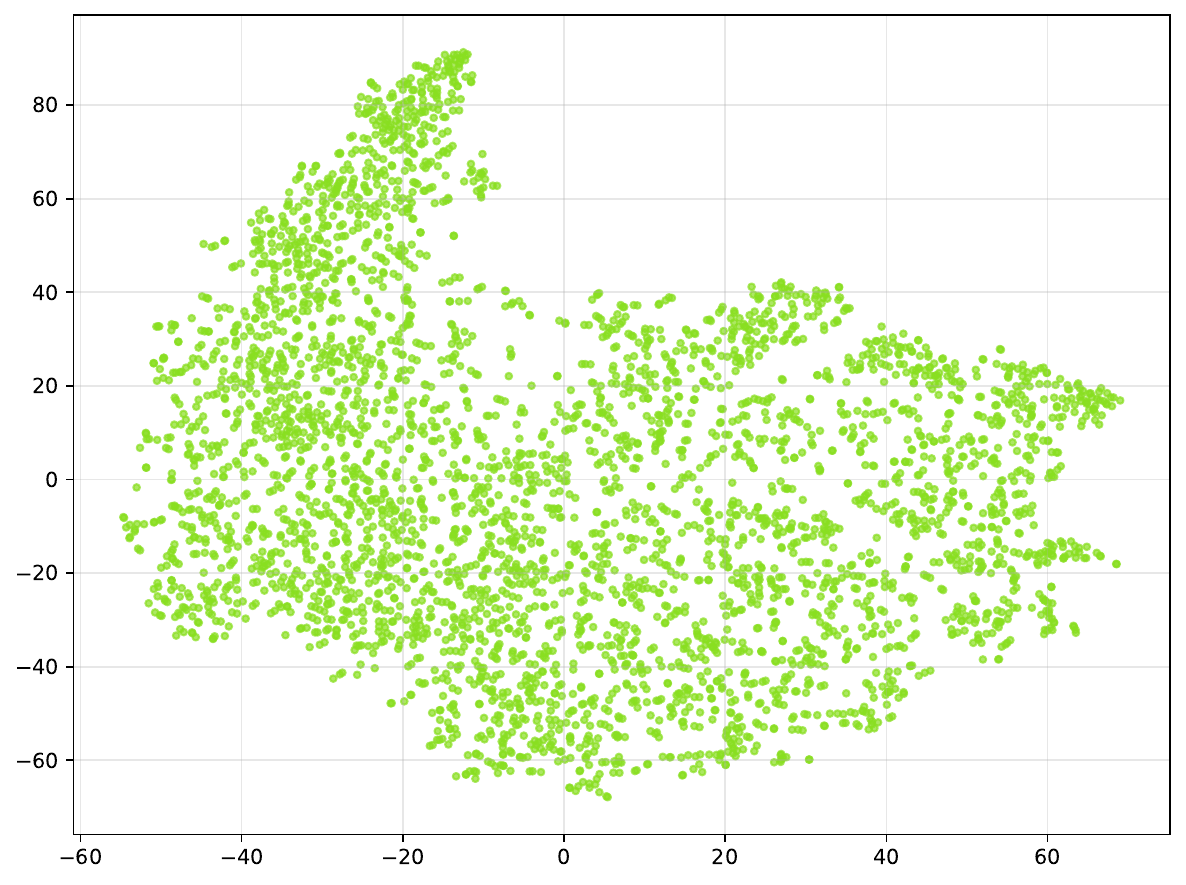}
        \subcaption{\footnotesize MLP$_3$ with KDEF}
    \end{minipage}
    \hspace{-0.5em} 
    
    \captionsetup{justification=raggedright}
    \caption{We visualize the t-SNE of the last layer of $\textbf{KDEF}_{\text{MLP}}$ on the ML-1M test set within a same batch.}
    \label{tsne}
    \vspace{-1em}
\end{figure}

Fig. \ref{tsne} uses t-SNE \cite{tsne} to visualize the feature distributions of the last layer of different sub-networks in $\textbf{KDEF}_{\text{MLP}}$. We observe that, under the influence of KD, the representations of the three MLPs are remarkably similar. We believe this is because the predictions of the same teacher model act as anchors for the sub-networks’ learning, leading to similar representations. DML learns from other student networks instead of relying solely on a fixed teacher, resulting in sub-networks having non-identical posterior distributions. This mechanism effectively introduces diversity among the sub-networks. Moreover, due to the introduction of the Examination Mechanism, the representations of sub-networks in KDEF become more distinct, alleviating the issue of information redundancy. This helps explain why KDEF achieves better performance.

\section{Conclusion and Future Work}
In this paper, we empirically revealed limitations and the cause of large ensemble networks in CTR prediction tasks and explored potential solutions from the perspectives of knowledge distillation and deep mutual learning. Accordingly, we proposed a novel model-agnostic and hyperparameter-free Knowledge-Driven Ensemble Framework (KDEF). Additionally, we designed a novel examination mechanism that balanced the weights of multiple losses to achieve tailored instruction from teachers to students and selective learning among students. Experiments on five real-world datasets demonstrated the effectiveness, compatibility, and flexibility of KDEF. For future work, we aimed to perform self-knowledge transfer or self-ensemble for sub-networks to further enhance the compatibility and effectiveness of the framework.

\section*{Acknowledgments}
This work is supported by the National Science Foundation of China (No. 62272001 and No. 62206002).



 
%

\bibliographystyle{IEEEtran}
\bibliography{cite}

\begin{thebibliography}{10}
\providecommand{\url}[1]{#1}
\csname url@samestyle\endcsname
\providecommand{\newblock}{\relax}
\providecommand{\bibinfo}[2]{#2}
\providecommand{\BIBentrySTDinterwordspacing}{\spaceskip=0pt\relax}
\providecommand{\BIBentryALTinterwordstretchfactor}{4}
\providecommand{\BIBentryALTinterwordspacing}{\spaceskip=\fontdimen2\font plus
\BIBentryALTinterwordstretchfactor\fontdimen3\font minus \fontdimen4\font\relax}
\providecommand{\BIBforeignlanguage}[2]{{%
\expandafter\ifx\csname l@#1\endcsname\relax
\typeout{** WARNING: IEEEtran.bst: No hyphenation pattern has been}%
\typeout{** loaded for the language `#1'. Using the pattern for}%
\typeout{** the default language instead.}%
\else
\language=\csname l@#1\endcsname
\fi
#2}}
\providecommand{\BIBdecl}{\relax}
\BIBdecl

\bibitem{Bars}
J.~Zhu, Q.~Dai, L.~Su, R.~Ma, J.~Liu, G.~Cai, X.~Xiao, and R.~Zhang, ``{Bars: Towards open benchmarking for recommender systems},'' in \emph{Proceedings of the 45th International ACM SIGIR Conference on Research and Development in Information Retrieval}, 2022, pp. 2912--2923.

\bibitem{TKDE1}
Y.~Li, X.~Guo, W.~Lin, M.~Zhong, Q.~Li, Z.~Liu, W.~Zhong, and Z.~Zhu, ``{Learning Dynamic User Interest Sequence in Knowledge Graphs for Click-through Rate Prediction},'' \emph{IEEE Transactions on Knowledge and Data Engineering}, vol.~35, no.~1, pp. 647--657, 2021.

\bibitem{TKDE2}
M.~Gao, J.-Y. Li, C.-H. Chen, Y.~Li, J.~Zhang, and Z.-H. Zhan, ``{Enhanced Multi-task Learning and Knowledge Graph-based Recommender System},'' \emph{IEEE Transactions on Knowledge and Data Engineering}, 2023.

\bibitem{aim}
C.~Zhu, B.~Chen, W.~Zhang, J.~Lai, R.~Tang, X.~He, Z.~Li, and Y.~Yu, ``{AIM: Automatic Interaction Machine for Click-Through Rate Prediction},'' vol.~35, no.~4, 2023, pp. 3389--3403.

\bibitem{TKDE3}
W.-J. Zhou, Y.~Zheng, Y.~Feng, Y.~Ye, R.~Xiao, L.~Chen, X.~Yang, and J.~Xiao, ``{ENCODE: Breaking the Trade-Off Between Performance and Efficiency in Long-Term User Behavior Modeling},'' \emph{IEEE Transactions on Knowledge and Data Engineering}, vol.~37, no.~1, pp. 265--277, 2025.

\bibitem{openbenchmark}
J.~Zhu, J.~Liu, S.~Yang, Q.~Zhang, and X.~He, ``{Open Benchmarking for Click-through Rate Prediction},'' in \emph{Proceedings of the 30th ACM International Conference on Information \& Knowledge Management}, 2021, pp. 2759--2769.

\bibitem{finalmlp}
K.~Mao, J.~Zhu, L.~Su, G.~Cai, Y.~Li, and Z.~Dong, ``{FinalMLP: An Enhanced Two-Stream MLP Model for CTR Prediction},'' \emph{Proceedings of the AAAI Conference on Artificial Intelligence, 37(4), 4552-4560.}, 2023.

\bibitem{dcnv2}
R.~Wang, R.~Shivanna, D.~Cheng, S.~Jain, D.~Lin, L.~Hong, and E.~Chi, ``{DCNv2: Improved Deep \& Cross Network and Practical Lessons for Web-scale Learning to Rank Systems},'' in \emph{Proceedings of the Web Conference 2021}, 2021, pp. 1785--1797.

\bibitem{ensemble_first}
X.~Ling, W.~Deng, C.~Gu, H.~Zhou, C.~Li, and F.~Sun, ``{Model Ensemble for Click Prediction in Bing Search Ads},'' in \emph{Proceedings of the 26th international conference on world wide web companion}, 2017, pp. 689--698.

\bibitem{GDCN}
F.~Wang, H.~Gu, D.~Li, T.~Lu, P.~Zhang, and N.~Gu, ``{Towards Deeper, Lighter and Interpretable Cross Network for CTR Prediction},'' in \emph{Proceedings of the 32nd ACM International Conference on Information and Knowledge Management}, 2023, pp. 2523--2533.

\bibitem{EulerNet}
Z.~Tian, T.~Bai, W.~X. Zhao, J.-R. Wen, and Z.~Cao, ``{EulerNet: Adaptive Feature Interaction Learning via Euler's Formula for CTR Prediction},'' in \emph{Proceedings of the 46th International ACM SIGIR Conference on Research and Development in Information Retrieval}, 2023, p. 1376–1385.

\bibitem{Wukong}
B.~Zhang, L.~Luo, Y.~Chen, J.~Nie, X.~Liu, S.~Li, Y.~Zhao, Y.~Hao, Y.~Yao, E.~D. Wen \emph{et~al.}, ``{Wukong: Towards a Scaling Law for Large-scale Recommendation},'' in \emph{Proceedings of the 41st International Conference on Machine Learning}, 2024, pp. 59\,421--59\,434.

\bibitem{xdeepfm}
J.~Lian, X.~Zhou, F.~Zhang, Z.~Chen, X.~Xie, and G.~Sun, ``{xDeepFM: Combining explicit and implicit feature interactions for recommender systems},'' in \emph{Proceedings of the 24th ACM SIGKDD International Conference on Knowledge Discovery \& Data Mining}, 2018, pp. 1754--1763.

\bibitem{autoint}
W.~Song, C.~Shi, Z.~Xiao, Z.~Duan, Y.~Xu, M.~Zhang, and J.~Tang, ``{AutoInt: Automatic feature interaction learning via self-attentive neural networks},'' in \emph{Proceedings of the 28th ACM International Conference on Information and Knowledge Management}, 2019, pp. 1161--1170.

\bibitem{ensemble}
T.~G. Dietterich \emph{et~al.}, ``{Ensemble learning},'' \emph{The handbook of brain theory and neural networks}, vol.~2, no.~1, pp. 110--125, 2002.

\bibitem{deepfm}
H.~Guo, R.~Tang, Y.~Ye, Z.~Li, and X.~He, ``{DeepFM: A Factorization-Machine Based Neural Network for CTR Prediction},'' in \emph{Proceedings of the 26th International Joint Conference on Artificial Intelligence}, ser. IJCAI'17.\hskip 1em plus 0.5em minus 0.4em\relax AAAI Press, 2017, p. 1725–1731.

\bibitem{KD4CTR}
J.~Zhu, J.~Liu, W.~Li, J.~Lai, X.~He, L.~Chen, and Z.~Zheng, ``{Ensembled CTR Prediction via Knowledge Distillation},'' in \emph{Proceedings of the 29th ACM International Conference on Information \& Knowledge Management}, 2020, pp. 2941--2958.

\bibitem{FINAL}
J.~Zhu, Q.~Jia, G.~Cai, Q.~Dai, J.~Li, Z.~Dong, R.~Tang, and R.~Zhang, ``{FINAL: Factorized interaction layer for CTR prediction},'' in \emph{Proceedings of the 46th International ACM SIGIR Conference on Research and Development in Information Retrieval}, 2023, pp. 2006--2010.

\bibitem{CETN}
H.~Li, L.~Sang, Y.~Zhang, X.~Zhang, and Y.~Zhang, ``{CETN: Contrast-enhanced Through Network for CTR Prediction},'' \emph{arXiv preprint arXiv:2312.09715}, 2023.

\bibitem{TF4CTR}
H.~Li, Y.~Zhang, Y.~Zhang, L.~Sang, and Y.~Yang, ``{TF4CTR: Twin Focus Framework for CTR Prediction via Adaptive Sample Differentiation},'' \emph{arXiv preprint arXiv:2405.03167}, 2024.

\bibitem{DHEN}
B.~Zhang, L.~Luo, X.~Liu, J.~Li, Z.~Chen, W.~Zhang, X.~Wei, Y.~Hao, M.~Tsang, W.~Wang \emph{et~al.}, ``{DHEN: A Deep and Hierarchical Ensemble Network for Large-scale Click-through Rate Prediction},'' \emph{arXiv preprint arXiv:2203.11014}, 2022.

\bibitem{PINTEREST}
S.~Malreddy, M.~Lawhon, U.~A. Nookala, A.~Mantha, and D.~D. Badani, ``Improving feature interactions at pinterest under industry constraints,'' \emph{arXiv preprint arXiv:2412.01985}, 2024.

\bibitem{tencent}
J.~Pan, W.~Xue, X.~Wang, H.~Yu, X.~Liu, S.~Quan, X.~Qiu, D.~Liu, L.~Xiao, and J.~Jiang, ``{Ads Recommendation in A Collapsed and Entangled World},'' in \emph{Proceedings of the 30th ACM SIGKDD Conference on Knowledge Discovery and Data Mining}, 2024, pp. 5566--5577.

\bibitem{MBCnet}
X.~Chen, Z.~Cheng, Y.~Pan, S.~Xiao, X.~Liu, J.~Lan, Q.~Liu, and I.~W. Tsang, ``{Branches, Assemble! Multi-Branch Cooperation Network for Large-Scale Click-Through Rate Prediction at Taobao},'' \emph{arXiv preprint arXiv:2411.13057}, 2024.

\bibitem{GPT3.0}
J.~Kaplan, S.~McCandlish, T.~Henighan, T.~B. Brown, B.~Chess, R.~Child, S.~Gray, A.~Radford, J.~Wu, and D.~Amodei, ``{Scaling Laws for Neural Language Models},'' \emph{arXiv preprint arXiv:2001.08361}, 2020.

\bibitem{KD_survey1}
J.~Gou, B.~Yu, S.~J. Maybank, and D.~Tao, ``{Knowledge Distillation: A Survey},'' \emph{International Journal of Computer Vision}, vol. 129, no.~6, pp. 1789--1819, 2021.

\bibitem{KD_survey2}
Z.~Li, P.~Xu, X.~Chang, L.~Yang, Y.~Zhang, L.~Yao, and X.~Chen, ``{When Object Detection Meets Knowledge Distillation: A Survey},'' \emph{IEEE Transactions on Pattern Analysis and Machine Intelligence}, vol.~45, no.~8, pp. 10\,555--10\,579, 2023.

\bibitem{KD}
G.~Hinton, ``{Distilling the Knowledge in a Neural Network},'' \emph{arXiv preprint arXiv:1503.02531}, 2015.

\bibitem{DML}
Y.~Zhang, T.~Xiang, T.~M. Hospedales, and H.~Lu, ``{Deep Mutual Learning},'' in \emph{Proceedings of the IEEE conference on computer vision and pattern recognition}, 2018, pp. 4320--4328.

\bibitem{KDCL}
Q.~Guo, X.~Wang, Y.~Wu, Z.~Yu, D.~Liang, X.~Hu, and P.~Luo, ``{Online Knowledge Distillation via Collaborative Learning},'' in \emph{Proceedings of the IEEE/CVF Conference on Computer Vision and Pattern Recognition}, 2020, pp. 11\,020--11\,029.

\bibitem{LR}
M.~Richardson, E.~Dominowska, and R.~Ragno, ``{Predicting Clicks: Estimating the Click-through Rate for New Ads},'' in \emph{Proceedings of the 16th International Conference on World Wide Web}, 2007, pp. 521--530.

\bibitem{FM}
S.~Rendle, ``{Factorization Machines},'' in \emph{2010 IEEE International Conference on Data Mining}.\hskip 1em plus 0.5em minus 0.4em\relax IEEE, 2010, pp. 995--1000.

\bibitem{DNN}
P.~Covington, J.~Adams, and E.~Sargin, ``{Deep Neural Networks for YouTube Recommendations},'' in \emph{Proceedings of the 10th ACM Conference on Recommender Systems}, 2016, pp. 191--198.

\bibitem{NFM}
X.~He and T.-S. Chua, ``{Neural Factorization Machines for Sparse Predictive Analytics},'' in \emph{Proceedings of the 40th International ACM SIGIR Conference on Research and Development in Information Retrieval}, 2017, pp. 355--364.

\bibitem{widedeep}
H.-T. Cheng, L.~Koc, J.~Harmsen, T.~Shaked, T.~Chandra, H.~Aradhye, G.~Anderson, G.~Corrado, W.~Chai, M.~Ispir \emph{et~al.}, ``{Wide \& Deep Learning for Recommender Systems},'' in \emph{Proceedings of the 1st Workshop on Deep Learning for Recommender Systems}, 2016, pp. 7--10.

\bibitem{adagin}
L.~Sang, H.~Li, Y.~Zhang, Y.~Zhang, and Y.~Yang, ``{AdaGIN: Adaptive Graph Interaction Network for Click-Through Rate Prediction},'' \emph{ACM Transactions on Information Systems}, 2024.

\bibitem{SimCEN}
H.~Li, L.~Sang, Y.~Zhang, and Y.~Zhang, ``{SimCEN: Simple Contrast-enhanced Network for CTR Prediction},'' in \emph{Proceedings of the 32th ACM International Conference on Multimedia}, 2024.

\bibitem{EDCN}
B.~Chen, Y.~Wang, Z.~Liu, R.~Tang, W.~Guo, H.~Zheng, W.~Yao, M.~Zhang, and X.~He, ``{Enhancing explicit and implicit feature interactions via information sharing for parallel deep CTR models},'' in \emph{Proceedings of the 30th ACM International Conference on Information \& Knowledge Management}, 2021, pp. 3757--3766.

\bibitem{FCN}
\BIBentryALTinterwordspacing
H.~Li, Y.~Zhang, Y.~Zhang, H.~Li, L.~Sang, and J.~Zhu, ``{FCN: Fusing Exponential and Linear Cross Network for Click-Through Rate Prediction},'' 2025. [Online]. Available: \url{https://arxiv.org/abs/2407.13349}
\BIBentrySTDinterwordspacing

\bibitem{TeacherKD1}
J.~H. Cho and B.~Hariharan, ``{On the Efficacy of Knowledge Distillation},'' in \emph{Proceedings of the IEEE/CVF international conference on computer vision}, 2019, pp. 4794--4802.

\bibitem{TeacherKD2}
Y.~Liu, W.~Zhang, and J.~Wang, ``{Adaptive Multi-teacher Multi-level Knowledge Distillation},'' \emph{Neurocomputing}, vol. 415, pp. 106--113, 2020.

\bibitem{DKD}
B.~Zhao, Q.~Cui, R.~Song, Y.~Qiu, and J.~Liang, ``{Decoupled knowledge distillation},'' in \emph{Proceedings of the IEEE/CVF Conference on computer vision and pattern recognition}, 2022, pp. 11\,953--11\,962.

\bibitem{KDNLP}
S.~Hahn and H.~Choi, ``{Self-Knowledge Distillation in Natural Language Processing},'' \emph{arXiv preprint arXiv:1908.01851}, 2019.

\bibitem{DAGFM}
Z.~Tian, T.~Bai, Z.~Zhang, Z.~Xu, K.~Lin, J.-R. Wen, and W.~X. Zhao, ``{Directed Acyclic Graph Factorization Machines for CTR Prediction via Knowledge Distillation},'' in \emph{Proceedings of the Sixteenth ACM International Conference on Web Search and Data Mining}, 2023, pp. 715--723.

\bibitem{BKD}
Y.~Deng, Y.~Chen, X.~Dong, L.~Pan, H.~Li, L.~Cheng, and L.~Mo, ``{BKD: A Bridge-based Knowledge Distillation Method for Click-Through Rate Prediction},'' in \emph{Proceedings of the 46th International ACM SIGIR Conference on Research and Development in Information Retrieval}, 2023, pp. 1859--1863.

\bibitem{PositionKD}
C.~Liu, Y.~Li, J.~Zhu, F.~Teng, X.~Zhao, C.~Peng, Z.~Lin, and J.~Shao, ``{Position Awareness Modeling with Knowledge Distillation for CTR Prediction},'' in \emph{Proceedings of the 16th ACM Conference on Recommender Systems}, 2022, pp. 562--566.

\bibitem{DML4CTR}
I.~C. Yilmaz and S.~Aldemir, ``{Mutual Learning for Finetuning Click-Through Rate Prediction Models},'' \emph{arXiv preprint arXiv:2406.12087}, 2024.

\bibitem{embedding_Collapse}
X.~Guo, J.~Pan, X.~Wang, B.~Chen, J.~Jiang, and M.~Long, ``{On the Embedding Collapse When Scaling up Recommendation Models},'' in \emph{Proceedings of the 41st International Conference on Machine Learning}, 2024, pp. 16\,891--16\,909.

\bibitem{Dimensional_Collapse}
\BIBentryALTinterwordspacing
L.~Jing, P.~Vincent, Y.~LeCun, and Y.~Tian, ``{Understanding Dimensional Collapse in Contrastive Self-supervised Learning},'' in \emph{International Conference on Learning Representations}, 2022. [Online]. Available: \url{https://openreview.net/forum?id=YevsQ05DEN7}
\BIBentrySTDinterwordspacing

\bibitem{feature_Collapse}
T.~Hua, W.~Wang, Z.~Xue, S.~Ren, Y.~Wang, and H.~Zhao, ``{On Feature Decorrelation in Self-supervised Learning},'' in \emph{Proceedings of the IEEE/CVF international conference on computer vision}, 2021, pp. 9598--9608.

\bibitem{SVD}
M.~E. Wall, A.~Rechtsteiner, and L.~M. Rocha, ``{Singular Value Decomposition and Principal Component Analysis},'' in \emph{A practical approach to microarray data analysis}.\hskip 1em plus 0.5em minus 0.4em\relax Springer, 2003, pp. 91--109.

\bibitem{dcn}
R.~Wang, B.~Fu, G.~Fu, and M.~Wang, ``{Deep \& cross network for ad click predictions},'' in \emph{Proceedings of the ADKDD'17}, 2017, pp. 1--7.

\bibitem{Emergent}
J.~Wei, Y.~Tay, R.~Bommasani, C.~Raffel, B.~Zoph, S.~Borgeaud, D.~Yogatama, M.~Bosma, D.~Zhou, D.~Metzler \emph{et~al.}, ``{Emergent Abilities of Large Language Models},'' \emph{arXiv preprint arXiv:2206.07682}, 2022.

\bibitem{Criteo}
kaggle, ``{The Criteo Dataset},'' \url{https://www.kaggle.com/c/criteo-display-adchallenge}, 2014.

\bibitem{dark_knowledge}
\BIBentryALTinterwordspacing
Z.~Allen-Zhu and Y.~Li, ``{Towards Understanding Ensemble, Knowledge Distillation and Self-Distillation in Deep Learning},'' in \emph{The Eleventh International Conference on Learning Representations}, 2023. [Online]. Available: \url{https://openreview.net/forum?id=Uuf2q9TfXGA}
\BIBentrySTDinterwordspacing

\bibitem{pnn2}
Y.~Qu, B.~Fang, W.~Zhang, R.~Tang, M.~Niu, H.~Guo, Y.~Yu, and X.~He, ``{Product-based neural networks for user response prediction over multi-field categorical data},'' \emph{ACM Transactions on Information Systems (TOIS)}, vol.~37, no.~1, pp. 1--35, 2018.

\bibitem{ComboFashion}
C.~Zhu, P.~Du, W.~Zhang, Y.~Yu, and Y.~Cao, ``{Combo-Fashion: Fashion Clothes Matching CTR Prediction with Item History},'' in \emph{Proceedings of the 28th ACM SIGKDD Conference on Knowledge Discovery and Data Mining}, 2022, pp. 4621--4629.

\bibitem{pnn1}
Y.~Qu, H.~Cai, K.~Ren, W.~Zhang, Y.~Yu, Y.~Wen, and J.~Wang, ``{Product-based Neural Networks for User Response Prediction},'' in \emph{2016 IEEE 16th International Conference on Data Mining (ICDM)}.\hskip 1em plus 0.5em minus 0.4em\relax IEEE, 2016, pp. 1149--1154.

\bibitem{AFN}
W.~Cheng, Y.~Shen, and L.~Huang, ``{Adaptive Factorization Network: Learning Adaptive-order Feature Interactions},'' in \emph{Proceedings of the AAAI Conference on Artificial Intelligence}, vol.~34, no.~04, 2020, pp. 3609--3616.

\bibitem{masknet}
Z.~Wang, Q.~She, and J.~Zhang, ``{MaskNet: Introducing Feature-wise Multiplication to CTR Ranking Models by Instance-guided Mask},'' \emph{arXiv preprint arXiv:2102.07619}, 2021.

\bibitem{CL4CTR}
F.~Wang, Y.~Wang, D.~Li, H.~Gu, T.~Lu, P.~Zhang, and N.~Gu, ``{CL4CTR: A Contrastive Learning Framework for CTR Prediction},'' in \emph{Proceedings of the Sixteenth ACM International Conference on Web Search and Data Mining}, 2023, pp. 805--813.

\bibitem{RFM}
Z.~Tian, Y.~Shi, X.~Wu, W.~X. Zhao, and J.-R. Wen, ``{Rotative Factorization Machines},'' in \emph{Proceedings of the 30th ACM SIGKDD Conference on Knowledge Discovery and Data Mining}, 2024, pp. 2912--2923.

\bibitem{PYTORCH}
A.~Paszke, S.~Gross, F.~Massa, A.~Lerer, J.~Bradbury, G.~Chanan, T.~Killeen, Z.~Lin, N.~Gimelshein, L.~Antiga \emph{et~al.}, ``{PyTorch: An imperative style, high-performance deep learning library},'' \emph{Advances in Neural Information Processing Systems}, vol.~32, 2019.

\bibitem{FuxiCTR}
Huawei, ``{An open-source CTR prediction library},'' \url{https://fuxictr.github.io}, 2021.

\bibitem{adam}
D.~P. Kingma and J.~Ba, ``{Adam: A Method for Stochastic Optimization},'' \emph{arXiv preprint arXiv:1412.6980}, 2014.

\bibitem{neuralvsmf}
S.~Rendle, W.~Krichene, L.~Zhang, and J.~Anderson, ``{Neural collaborative filtering vs. matrix factorization revisited},'' in \emph{Proceedings of the 14th ACM Conference on Recommender Systems}, 2020, pp. 240--248.

\bibitem{tsne}
L.~Van~der Maaten and G.~Hinton, ``{Visualizing data using t-SNE.}'' \emph{Journal of Machine Learning Research}, vol.~9, no.~11, 2008.

\end{thebibliography}

\begin{IEEEbiography}
[{\includegraphics[width=1in,height=1.25in,clip,keepaspectratio]{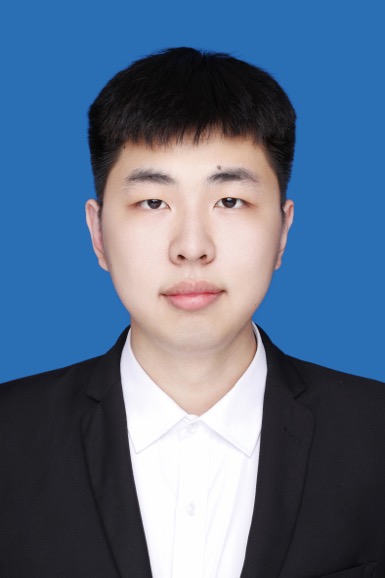}}]{Honghao Li}
received the Bachelor degree in Computer engineering and Technology from Bengbu University, Bengbu, China, in 2022. He is currently pursuing a Ph.D. degree at Anhui University's School of Computer Science and Technology. His current research interests include CTR prediction, service computing, and recommender systems.
\vspace{-1em}
\end{IEEEbiography}

\begin{IEEEbiography}[{\includegraphics[width=1in,height=1.25in,clip,keepaspectratio]{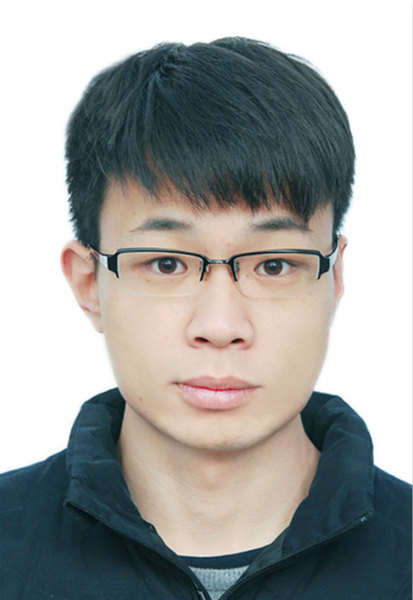}}]{Lei Sang}
received the Ph.D. degree from the Faculty of Engineering and Information Technology, University of Technology Sydney, Sydney, Australia, in 2021.
He is currently a Lecturer with the School of Computer Science and Technology, Anhui University, Anhui, China. His current research interests include natural language processing, data mining, and recommendation systems.
\vspace{-1em}
\end{IEEEbiography}

\begin{IEEEbiography}[{\includegraphics[width=1in,height=1.25in,clip,keepaspectratio]{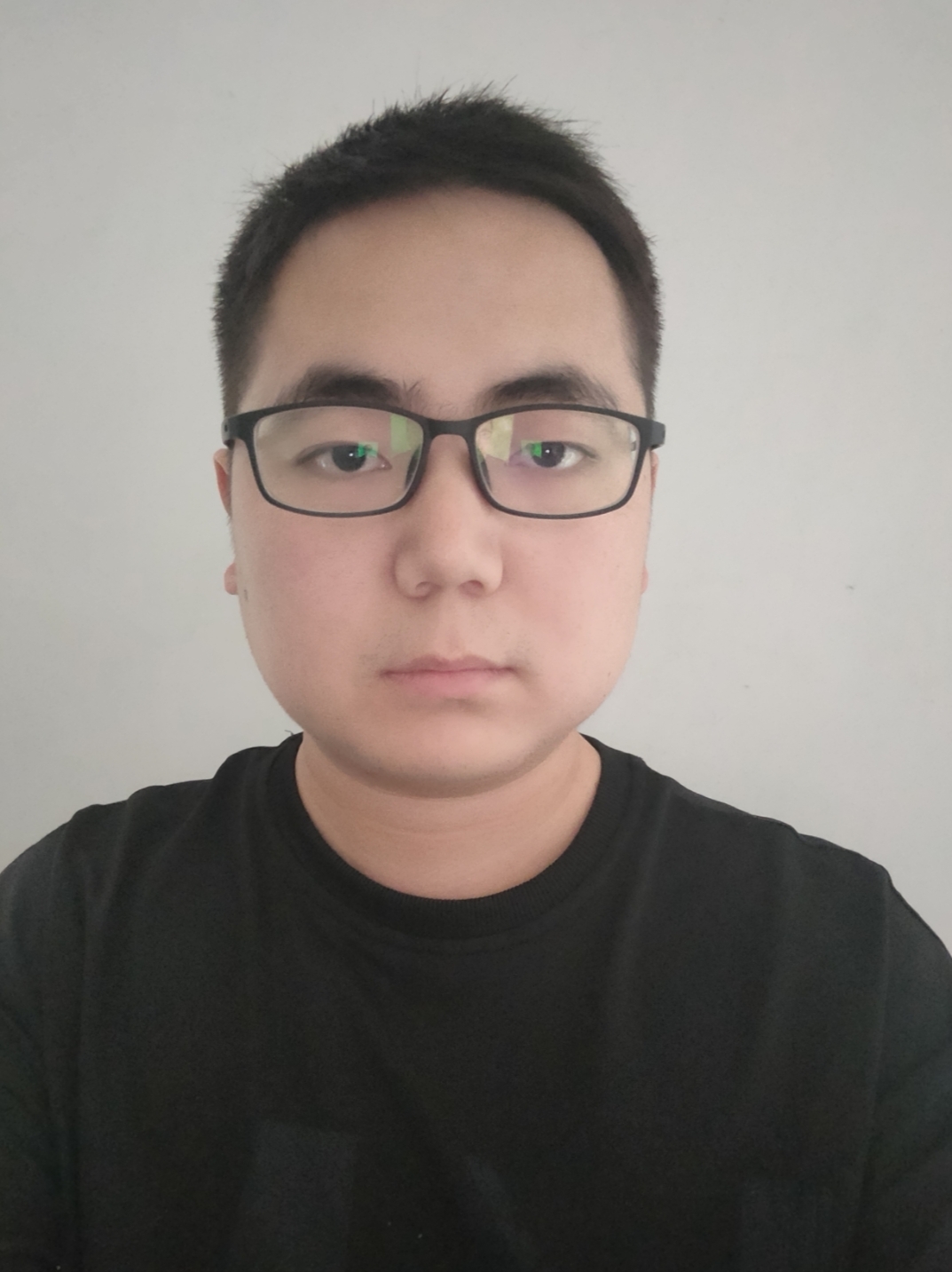}}]{Yi Zhang}
received the Bachelor  degree in Computer Science and Technology from Anhui University,
Hefei, China, in 2020, where he is currently pursuing a Ph.D. degree. He has publications in several top conferences and journals, including IEEE TKDE, IEEE TSMC, IEEE TBD, ACM TOIS, and ACM SIGIR, etc. His current research interests include graph neural network, personalized recommender systems, and service computing.
\vspace{-1em}
\end{IEEEbiography}

\begin{IEEEbiography}[{\includegraphics[width=1in,height=1.25in,clip,keepaspectratio]{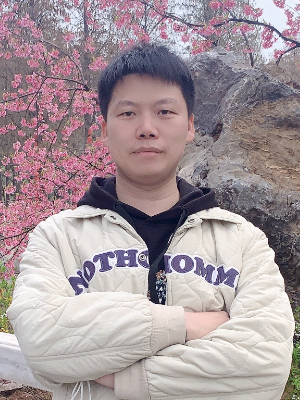}}]{Guangming Cui}
received his Master's degree from Anhui University, China, in 2018 and his PhD degree from Swinburne University of Technology, Australia, in 2022, in computer science. Currently, he is an associate professor at Nanjing University of Information Science \& Technology, China. He has published more than 30 peer-reviewed articles in international journals and conferences, including the IEEE TMC, IEEE TPDS, IEEE TSC, IEEE TDSC, JSAC, ICWS, ICSOC, etc. His research interests include edge computing, service computing, mobile computing, and software engineering.
\vspace{-1em}
\end{IEEEbiography}

\begin{IEEEbiography}[{\includegraphics[width=1in,height=1.25in,clip,keepaspectratio]{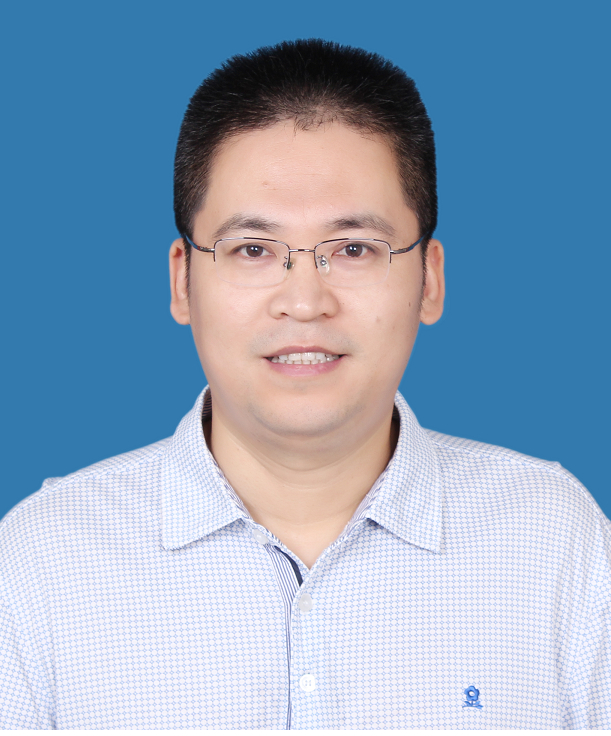}}]{Yiwen Zhang}
received the Ph.D. degree in management science and engineering from Hefei University of Technology, in 2013. He is currently a full professor with the School of Computer Science and Technology, Anhui University. He has published more than 70 papers in highly regarded conferences and journals, including IEEE TKDE, IEEE TMC, IEEE TSC, ACM TOIS, IEEE TPDS, IEEE TNNLS, ACM TKDD, SIGIR, ICSOC, ICWS, etc. His research interests include service computing, cloud computing, and big data analytics. Please see more information in our website \url{http://bigdata.ahu.edu.cn/}.
\vspace{-1em}
\end{IEEEbiography}

\end{document}